\documentstyle[aps,epsf,cite,array]{revtex}

\def\nn{\nonumber}

\def\la{\langle}
\def\ra{\rangle}

\def\l{\left}
\def\r{\right}
\def\nn{\nonumber}

\def\ord#1{{\mathcal{O}}\l(#1\r)}
\def\cO{{\mathcal{O}}}

\def\beq{\begin{equation}}
\def\eeq{\end{equation}}
\def\bea{\begin{eqnarray}}
\def\eea{\end{eqnarray}}

\def\eq#1{Eq.\ {(\ref{#1})}}
\def\eqs#1#2{Eqs.\ {(\ref{#1})} and {(\ref{#2})}}
\def\eqss#1#2#3{Eqs.\ {(\ref{#1})}, {(\ref{#2})} and {(\ref{#3})}}
\def\tab#1{Table {\ref{#1}}}
\def\tabss#1#2#3{Tables {\ref{#1}}, {\ref{#2}} and {\ref{#3}}}
\def\sec#1{Section {\ref{#1}}}
\def\secs#1#2{Sections {\ref{#1}} and {\ref{#2}}}
\def\fig#1{Figure {\ref{#1}}}
\def\figs#1#2{Figures {\ref{#1}} and {\ref{#2}}}
\def\figss#1#2#3{Figures {\ref{#1}}, {\ref{#2}} and {\ref{#3}}}
\def\figsss#1#2#3#4{Figures {\ref{#1}}, {\ref{#2}}, {\ref{#3}} and {\ref{#4}}}
\def\err#1#2{\raisebox{0.3ex}{\footnotesize $\begin{array}{l}{ +
#1}\\ { - #2}\end{array}$}}

\def\mev{\mbox{ MeV}}
\def\gev{\mbox{ GeV}}
\def\msbar{\overline{\mbox{MS}}}
\def\msbartiny{\overline{\mbox{\tiny MS}}}
\def\hqettiny{\mbox{\tiny HQET}}
\def\ln{{\mathrm{ln}}}
\def\exp{{\mathrm{exp}}}
\def\tr{{\mathrm{Tr}}}

\def\lat{{\mathrm{lat}}}
\def\upl{U_{\mathrm{p}}}
\def\nlo{{\mathrm{nlo}}}
\def\csw{c_{\mbox{\tiny SW}}}

\begin{document}

\draft


\title{Standard model matrix elements for neutral $B$-meson mixing
and associated decay constants}

\author{Laurent Lellouch} 
\address{ LAPTH, Chemin de Bellevue, B.P. 110, F-74941 Annecy-le-Vieux
Cedex, France\\ and \\ Centre de Physique Th\'eorique, Case 907, CNRS
Luminy, F-13288 Marseille Cedex 9, France}

\author{C.-J. David Lin}
\address{ Department of Physics and Astronomy, 
      The University of Southampton, Southampton SO17 1BJ, England}

\author{(UKQCD Collaboration)}

\maketitle

\begin{abstract}
We present results of quenched lattice calculations of the
matrix elements relevant for $B_d{-}\bar B_d$ and $B_s{-}\bar B_s$ mixing
in the standard model. Results for the corresponding $SU(3)$-breaking
ratios, which can be used to constrain or determine $|V_{td}|$, are
also given.  The calculations are performed at two values of the
lattice spacing, corresponding to $\beta = 6.0$ and $\beta = 6.2$,
with quarks described by a mean-field-improved Sheikholeslami-Wohlert
action.  As a by-product, we obtain the leptonic decay constants of
$B$ and $D$ mesons.  We also present matrix elements relevant for
$D^0{-}\bar D^0$ mixing. Our results are summarized in the
Introduction.

\end{abstract}


\section{Introduction}
\label{sec:intro}

The study of $B_d{-}\bar B_d$ oscillations enables measurement of the
magnitude of the poorly known Cabibbo-Kobayashi-Maskawa (CKM) matrix
element, $V_{td}$, and thus the determination of one of the sides of
the unitarity triangle. The frequency of these oscillations is given
by the mass difference,
\begin{equation}
\Delta m_d \equiv M^{H}_{B_d}-M^{L}_{B_d} \ ,
\label{eq:dmd}
\end{equation}
where $M^{H}_{B_d}$ and $M^{L}_{B_d}$ are the heavy and light mass
eigenvalues of the $B_d{-}\bar B_d$ system.  $\Delta m_d$ is
experimentally measurable from tagged $B_d$ meson
samples.~\footnote{For a recent experimental review see, for instance,
\cite{Stocchi:2000ps} or \cite{Daoudi:1999pj}.} It is also calculable
in the standard model.  Keeping only dimension six operators after
an operator product expansion in which the top quark and $W$
boson are integrated out, the standard model prediction for $\Delta
m_d$ is, to next to leading order (NLO)
\cite{Buras:1990fn,Buchalla:1996vs}:
\beq
\Delta m_d = \frac{G_F^2}{8\pi^2}\, M_W^2\, |V_{td}
V_{tb}^*|^2\, 
\eta_B S_0(x_t)C_B(\mu)\, 
\frac{|\la \bar B_d|\cO^{\Delta B=2}_d(\mu)|B_d\ra|}{2M_{B_d}}
\ ,
\label{eq:deltamd}
\eeq
where $x_{t} = m_t^2/M^{2}_{W}$, $S_0(x_{t})\simeq 0.784\,x_t^{0.76}$
(to better than 1\%) is the relevant Inami-Lim function
\cite{InamiLim}, $\mu$ the renormalization scale, $\cO^{\Delta
B=2}_d$ the four-quark operator $\l[\bar
b\gamma^\mu(1-\gamma^5)d\r]$ $\l[\bar b\gamma_\mu(1-\gamma^5)d\r]$ and
$\eta_B=0.55$ and $C_B(\mu)$, short-distance coefficients.  The
renormalization-scale dependence of $C_B(\mu)$ and of the hadronic
matrix element cancel such that $\Delta m_d$ is $\mu$--independent,
to the order in perturbation theory at which $C_B(\mu)$ is
calculated. In the naive dimensional regularization modified 
minimal subtraction (NDR-$\msbar$) scheme assumed here,
\beq
C_B(\mu)=[\alpha_s(\mu)]^{-6/23}\l[1+\frac{\alpha_s(\mu)}{4\pi}J_5\r],
\quad J_5=\frac{5165}{3174}\ .
\label{eq:cbdef}
\eeq
Since $|V_{tb}|$ is equal to unity to very good accuracy, a
measurement of $\Delta m_d$ clearly enables the determination of
$|V_{td}|$. The accuracy of this determination is limited, at present,
by the theoretical uncertainty in the calculation of the
nonperturbative, strong-interaction effects in the matrix element
$\la \bar B_d|\cO^{\Delta B=2}_d(\mu)| B_d\ra$.  

\medskip

An alternative approach, in which many theoretical uncertainties
cancel, is to consider the ratio, $\Delta m_s/\Delta m_d$, where
$\Delta m_s$ is the mass difference in the neutral $B_s{-}\bar B_s$
system. In the standard model,
\beq
\frac{\Delta m_s}{\Delta m_d}=
\l|\frac{V_{ts}}{V_{td}}\r|^2\frac{M_{B_s}}{M_{B_d}}\xi^2=
\l|\frac{V_{ts}}{V_{td}}\r|^2 \frac{M_{B_d}}{M_{B_s}}r_{sd}
\equiv\l|\frac{V_{ts}}{V_{td}}\r|^2\ 
\frac{M_{B_d}}{M_{B_s}}
\l|\frac{\la \bar B_s|\cO^{\Delta B=2}_s|B_s\ra}
{\la\bar B_d|\cO^{\Delta B=2}_d|B_d\ra}\r|
\ , 
\label{eq:deltamsovermd}
\eeq
where $\cO^{\Delta B=2}_s$ is the same operator as
$\cO^{\Delta B=2}_d$, with $d$ replaced by $s$, and where we
have omitted the renormalization-scale dependence, as it cancels in the
ratio. Because the unitarity of the CKM matrix implies
$|V_{ts}|{\simeq} |V_{cb}|$ to a few percent and a clean extraction of
$|V_{cb}|$ can be achieved by analyzing semileptonic $B$ decays
\cite{Poling:1999aa}, a measurement of $\Delta m_s/\Delta m_d$ yields
a determination of $|V_{td}|$.  The high frequency of $B_s{-}\bar B_s$
oscillations makes this a challenging measurement.  Nevertheless, the
experimental lower bounds obtained on $\Delta m_s$
\cite{Stocchi:2000ps,Daoudi:1999pj} already yield interesting
constraints on the unitarity triangle
\cite{Buras:1997th,Mele:1998bf,Ali:1999we,Ciuchini:1999xh,
Plaszczynski:1999xs,Caravaglios:2000an,Stocchi:2000ps}.

\medskip

The matrix elements that appear in \eq{eq:deltamsovermd} are
traditionally parametrized as
\beq
\la\bar B_q|\cO^{\Delta B=2}_q(\mu)|B_q\ra = 
\frac{8}{3} M_{B_q}^2 f_{B_q}^2 B_{B_q}(\mu)
\ ,\label{eq:bparamdef}
\eeq
where $q=d$ or $s$ and where the parameter, $B_{B_q}(\mu)$, measures
deviations from vacuum saturation, corresponding to
$B_{B_q}(\mu)=1$. Here, $f_{B_q}$ is the decay constant
defined by
\beq
\la 0| \bar b\gamma_\mu\gamma^5 q|B_q(\vec{p})\ra = 
ip_\mu f_{B_q}
\ .\label{eq:fdef}
\eeq

\medskip

One also usually introduces a renormalization-group invariant and
scheme-independent parameter, $\hat B_{B_q}$, which to NLO in QCD is
given by
\beq
\hat B_{B_q}^\nlo=C_B(\mu) B_{B_q}(\mu)\ ,
\label{eq:bnlodef}
\eeq
where $C_B(\mu)$ is given by \eq{eq:cbdef} as long as $B_{B_q}(\mu)$
is computed in the NDR-$\msbar$ scheme with five active quarks. For
consistency with the value of $\eta_B$ given after \eq{eq:deltamd}, 
$\alpha_s$ should be taken to have its two-loop value with
$\Lambda_{\msbartiny}^{(5)}=225\mev$.

\medskip

In the present paper we report on high statistics, quenched lattice
QCD calculations of matrix elements and the corresponding $SU(3)$-breaking
ratios relevant for neutral $B$-meson mixing.  We obtain the
$SU(3)$-breaking ratio $r_{sd}$ in two ways: (1) by calculating
$B_{B_s}/B_{B_d}$ and $f_{B_s}/f_{B_d}$ and combining these two ratios
with the experimental mass ratio, $M_{B_s}/M_{B_d}$ (``indirect''
method); (2) by calculating the matrix elements, ${\la\bar
B_{d,s}|\cO^{\Delta B=2}_{d,s}|B_{d,s}\ra}$, directly and
taking their ratio (``direct'' method), as suggested in Ref. 
\cite{Bernard:1998dg} \footnote{Our method differs slightly from
that of Ref. \cite{Bernard:1998dg} in that we actually calculate
$(\la\bar{B}_{s}|{\mathcal{O}}^{\Delta B=2}_{s}|B_{s}\ra/
 \la\bar{B}_{d}|{\mathcal{O}}^{\Delta B=2}_{d}|B_{d}\ra)\times 
 (M_{B_{d}}/M_{B_{s}})$ and multiply by the experimentally measured
value of $M_{B_{s}}/M_{B_{sd}}$.}.  
They mainly differ in the required light- and heavy-quark-mass
interpolations and extrapolations, since for the ``direct'' method it
is the matrix element and corresponding ratio that are interpolated
and extrapolated, while for the ``indirect'' method it is the
$B$-parameters, decay constants, and corresponding ratios.

\medskip

As described in more detail in \sec{sec:simdtl}, these calculations
are performed at two values of the lattice spacing, $a$ ($\sim
(2.0\gev)^{-1}$ and $\sim (2.7\gev)^{-1}$), with relativistic Wilson
fermions. In order to keep discretization errors in check, the lattice
calculation is performed with heavy quarks whose masses are around
that of the charm and the results are extrapolated to the mass of the
$b$. Even in the charm sector, however, quarks have compton
wavelengths that are not much larger than our lattice spacings, and it
is important to reduce discretization errors as much as possible. We
attempt to do so by describing quarks with
mean-field-improved~\cite{Lepage:1993xa}, Sheikholeslami-Wohlert (SW)
actions \cite{Sheikholeslami:1985ij}.  When combined with improved
operators, these actions lead to discretization errors, which are
formally smaller than those generated with an unimproved Wilson action
($\ord{\alpha_sa}$ instead of $\ord{a}$), and which may be numerically
smaller than those brought about by a tree-level-improved SW action.
It is important to note that, as far as four-quark operators are
concerned, nonperturbative $\ord{a}$-improvement has not yet been
undertaken, and all lattice calculations of $B_{d,s}{-}\bar B_{d,s}$
mixing matrix elements have, as we do, $\ord{\alpha_sa}$
discretization errors, or worse, $\ord{a}$ errors. It is also
important to remember that $\ord{a^2}$ errors can be significant in
the presence of heavy quarks in a relativistic approach.

\medskip

Alternatively, one could take an effective theory approach and work
with static, nonrelativistic QCD (NRQCD)
or Fermilab quarks. In these approaches, the matrix
elements are expanded in inverse powers of the heavy-quark mass to
remove it from the long-distance dynamics. One important advantage is
that discretization errors are no longer enhanced by this mass. An
accurate description of the physics of the $b$ quark, however,
requires one to consider corrections in inverse powers of $m_b$ and
the calculation of these corrections is made difficult by
contributions proportional to inverse powers of $a$. The effective
theory and relativistic approaches should thus be viewed as
complementary.

\medskip

An additional feature of our calculation is that we extrapolate
$SU(3)$-breaking ratios in the heavy-quark mass directly instead of first
extrapolating numerator and denominator and then taking their ratio.
The heavy-quark-mass dependence will cancel partially between
numerator and denominator and therefore make the extrapolation more
reliable. This approach turns out to be particularly fruitful for the
determination of $f_{B_s}/f_{B}$, where the statistical error is
significantly reduced by a direct extrapolation of the ratio.

\medskip

Our main results are~\footnote{The results of preliminary analyses on
the same lattices were presented in Refs.
\cite{Lellouch:1998xk,Lellouch:1999ir,
Lellouch:1999ym}.}$^,$\footnote{We take $M_{B_d}=5279\mev$,
$M_{B_s}=5375\mev$, $M_D=1864\mev$ and $M_{D_s}=1969\mev$.}
$$
\xi = \frac{f_{B_s}\sqrt{B_{B_s}}}
{f_{B_d}\sqrt{B_{B_d}}}=1.15(2)\err{4}{2}\ ,
\quad\quad
r_{sd} = \l(\frac{M_{B_s}}{M_{B_d}}\r)^2\xi^2 = 1.38(6)\err{10}{6}
\ ,
$$
\beq
\begin{array}{rclrcl}
f_{B_d}\sqrt{\hat B_{B_d}^\nlo} &=& 210(21)\err{27}{26}\mev\ ,
\quad\quad
&\frac{f_{B_d}}{f_{D_s}}\sqrt{\hat B_{B_d}^\nlo} &=& 0.89(7)\err{6}{7}
\ ,\\
f_{B_s}\sqrt{\hat B_{B_s}^\nlo} &=& 241(14)\err{30}{27}\mev\ ,
\quad\quad
&\frac{f_{B_s}}{f_{D_s}}\sqrt{\hat B_{B_s}^\nlo} &=& 1.02(4)\err{6}{7}
\ ,
\label{eq:mainresults}
\end{array}
\eeq
\bea
&\begin{array}{rclrcl}
 B_{B_d}(M_{B}) &=& 0.91(4)\err{4}{0},\quad\quad & \hat B_{B_d}^\nlo &=& 
 1.40(5)\err{6}{1}\ ,\\
 B_{B_s}(M_{B}) &=& 0.90(2)\err{3}{0},\quad\quad &\hat B_{B_s}^\nlo &=&
 1.38(3)\err{5}{0}\ ,
\end{array}&
\nonumber\\
& B_{B_s}/B_{B_d} = 0.98(2)\err{0}{2}\ ,&\nonumber
\eea
where the first error is statistical and the second corresponds to the
systematic uncertainties added in quadrature. In quantities involving
ratios of $B$ parameters, the renormali\-zation-scale dependence is
not specified, as it cancels. We consider a wide array of systematic
uncertainties, as discussed in \sec{sec:results}. We normalize
dimensionful quantities, involving decay constants, by $f_{D_s}$
because some systematic (and statistical) uncertainties, including
possibly those associated with quenching, partially cancel in the
ratio. The original quantities can then be recovered by using the
experimental measurement of
$f_{D_s}$~\footnote{Ref. \cite{Parodi:1999nr} gives
$f_{D_s}=241(32)\mev$ as a summary number, which is in good agreement
with our determination of this quantity (see \eq{eq:decayconstants}
below). However, recent determinations appear to yield larger values,
albeit with large uncertainties: $f_{D_s}=285(20)(40)\mev$
\cite{AlephfDs2000}, $f_{D_s}=323(44)(36)\mev$
\cite{Alexandrov:2000ns} and $f_{D_s}=280(19)(44)\mev$
\cite{Chadha:1998zh}.}.

\medskip

The results of \eq{eq:mainresults} will be compared, in \sec{final},
to earlier calculations of some or all of these quantities performed
with propagating heavy quarks
\cite{Bernard:1998dg,Aoki:1995jq,Becirevic:2000nv} and with
non-relativistic quarks
\cite{Hashimoto:2000eh,Yamada:2000ym}. The comparison of the
$B$ parameters with results obtained using static heavy quarks
\cite{Ewing:1996ih,Gimenez:1997sk,Christensen:1997sj,Gimenez:1998mw}
will be addressed elsewhere \cite{bstatinprep}.

\medskip

Because decay constants and the corresponding matrix elements are
necessary for obtaining the results of \eq{eq:mainresults}, we also
have results for these decay constants. We find
\beq
\begin{array}{rclcrcl}
f_{B}&=& 177(17)\err{22}{22}\mev\ ,&\mbox{\hspace{3cm}}& & & \\
f_{B_s}&=&204(12)\err{24}{23}\mev\ , & & f_D&=&
210(10)\err{17}{16}\mev\ ,\\
\frac{f_{B_s}}{f_{B}}&=&1.15(2)\err{4}{2}\ , & & 
f_{D_s}&=&236(8)\err{17}{14}\mev\ ,\\
\frac{f_{B}}{f_{D_s}}&=&0.71(6)\err{4}{5}\ , & & \frac{f_{D_s}}{f_D}&=&
1.13(2)\err{4}{2}\ ,\\
\frac{f_{B_s}}{f_{D_s}}&=& 0.82(3)\err{4}{5}\ ,& & & &
\end{array}
\label{eq:decayconstants}
\eeq
where the first error is statistical and the second systematic, as
discussed in \sec{sec:results}. A comparison with recent quenched
results \cite{Allton:1997tv,El-Khadra:1998hq,AliKhan:1998df,Bernard:1998xi,
Ishikawa:1999xu,Bowler:2000xw,Khan:2000eg,Bernard:2000unq} 
will be made in \sec{final} and a discussion of unquenched results
\cite{Bernard:1999nv,Collins:1999ff,Khan:2000eg,Bernard:2000unq} will
be undertaken in \sec{sec:quench}. Note that results for $f_B$
can be combined with the measurement of the branching ratio for the
rare decay $B^+\to\tau^+\nu_\tau$, when it becomes available, to yield
a clean determination of $|V_{ub}|$.~\footnote{For recent reviews of
lattice calculations of the quantities presented in
\eqs{eq:mainresults}{eq:decayconstants}, please see
\cite{Wittig:1997bf,Flynn:1997ca,Draper:1998ms,Sharpe98,
Lellouch:1999ir,Hashimoto:1999bk,Lubicz:2000dj}.}

\medskip

While short-distance $D^0{-}\bar D^0$ mixing is highly suppressed in
the standard model \cite{Datta:1985jx}, it can be enhanced in
supersymmetric extensions
\cite{Barbieri:1982gn,Ellis:1982ts,Gabbiani:1989rb,Nir:1993mx,Leurer:1994gy},
above even the long-distance contributions discussed in Ref. 
\cite{Donoghue:1986hh}.  Thus, we give the $B$ parameter and
decay-constant combinations relevant for the matrix element of the
left-left, $\Delta C=2$ operator, which is one of the operators
that can contribute in supersymmetric extensions:
\beq
\begin{array}{c}
B_{D}(M_{D}) = 0.82(3)\err{4}{1}\ , \quad\quad
\hat{B}^{\mathrm{nlo}}_{D} = 1.12(4)\err{5}{1}\ ,\\
f_D\sqrt{\hat{B}^{\mathrm{nlo}}_{D}} = 222(10)\err{20}{16}\mev\ ,\quad\quad
\frac{f_D}{f_{D_{s}}}\sqrt{\hat{B}^{\mathrm{nlo}}_{D}} = 
0.94(3)\err{2}{3}\ ,
\end{array}
\label{eq:ddbar}
\eeq
where $\hat{B}^{\mathrm{nlo}}_{D}$ is obtained by multiplying
$B_{D}(M_{D})$ by $C_{D}(M_{D})$, with
\beq
 C_{D}(\mu) = \left [ \alpha_{s}(\mu)\right ]^{-6/25}
 \left [ 1 + \frac{\alpha_{s}(\mu)}{4\pi}J_{4}\right ] ,\mbox{ }\mbox{ }
 J_{4}=\frac{6719}{3750} ,
\eeq
where the two-loop $\alpha_{s}(\mu)$ is evaluated with 
$\Lambda^{(4)}_{\msbartiny}=350$ MeV.

\section{Simulation details}
\label{sec:simdtl}

Our results are based on quenched, $SU(3)$ gauge configurations,
calculated on a $24^3\times 48$ lattice at $\beta=6.2$ and a
$16^3\times 48$ lattice at $\beta=6.0$. The configurations are
generated using the hybrid over-relaxed algorithm described
in Ref. \cite{Allton:1993ue}. The parameters of the simulations are
summarized in \tab{tab:simparam}.

\medskip

We describe quarks with the Sheikholeslami-Wohlert (SW) 
action \cite{Sheikholeslami:1985ij} \footnote{Here and below
$\sigma_{\mu\nu} = \frac{i}{2} \l [ \gamma_{\mu}, \gamma_{\nu}\r ]$,
where $\gamma_{\mu}$ are the usual set of Euclidean $\gamma$-matrices
with $\l \{ \gamma_{\mu}, \gamma_{\nu}\r \} = 2\delta_{\mu\nu}$.}
\beq
  S_{\mathrm{F}}^{\mathrm{SW}} = S_{\mathrm{F}}^{\mathrm{W}} - 
 ig_0\,\csw\frac{\kappa_\psi}{2}\sum_{x,\mu,\nu}\,
             \bar \psi(x)\,P_{\mu\nu}(x)\sigma_{\mu\nu}\,\psi(x)\,
\ ,
\label{eq:sw}
\eeq
where $S_{\mathrm{F}}^{\mathrm{W}}$ is the standard Wilson action, $g_0$
the bare gauge coupling, $P_{\mu\nu}$ a lattice definition of the
field strength tensor, $\kappa_\psi$ the appropriate quark hopping
parameter and $\csw$, the so-called clover coefficient. Here, $\psi$
stands for both light ($q$) and heavy ($Q$) quarks.  While with the
Wilson action ($\csw=0$) spectral quantities suffer from
discretization errors of $\cO(a)$, the tree-level value,
$\csw=1$, guarantees that these errors are reduced to
$\cO(\alpha_s a)$
\cite{Sheikholeslami:1985ij,Heatlie:1991kg}.  In the present paper we
work with a mean-field estimate of the clover
coefficient~\cite{Lepage:1993xa,Shanahan:1997pk}, $\csw= 1/u_0^3$ with
$u_0=\la \frac{1}{3}\tr\upl \ra^{\frac{1}{4}}$.  Since this estimate
accounts for large tadpole contributions and is closer to the
nonperturbative value of the clover coefficient \cite{Luscher:1997ug}
which removes $\cO(a)$ errors to all orders in $\alpha_s$, our
discretization errors may be numerically smaller than for $\csw=1$. It
is important to note, however, that in the presence of heavy quarks
with masses $am_Q\sim 0.3$ or more, discretization errors of
$\ord{(am_Q)^2}$ may be comparable to those of $\ord{\alpha_sam_Q}$.

\medskip

In order to improve matrix elements up to 
$\ord{\alpha_s a}$, we must further ``rotate'' the quark 
fields \cite{Heatlie:1991kg}:
\beq
\psi \to \l(1-\frac{a}{2}\l(z\gamma\cdot\vec{D}-(1-z)m_\psi\r)\r) \psi
\ ,
\label{eq:rot}
\eeq
where $\vec{D}_\mu$ is the symmetric covariant derivative, $m_\psi$ the
bare quark mass to be defined later and $z$ is a real parameter which
can have any value between 0 and 1. For heavy quarks, a large source
of discretization errors is the mismatch in the normalization of
tree-level, zero-momentum, continuum and lattice quark
propagators. \eq{eq:rot} with $z=0$ corrects this mismatch at
$\ord{a}$.  Since we can compensate this mismatch completely by
implementing the El-Khadra-Kronfeld-Mackenzie (EKM) 
normalization \cite{El-Khadra:1997mp},
\beq
\psi \to \sqrt{1+am_\psi}\ \psi
\ ,
\label{eq:EKLM}
\eeq
we choose the latter instead of \eq{eq:rot}.

\medskip

At both values of the lattice spacing, we work with several values of
the heavy-quark hopping parameter straddling the charm. This enables
us to extrapolate our results in heavy-quark mass from the charm
sector, where discretization errors appear to be only a fraction of
the result, to the bottom sector, where these errors would be very
large were we to perform the simulation directly with such quarks. We
also consider several values of the light-quark hopping parameter
around that of the strange. Then, we interpolate our results in
the light-quark mass to the strange and extrapolate them to the chiral
limit. The values of the hopping parameters used in our paper are given
in \tab{tab:qmasses}. For completeness, the masses of the
corresponding light-light and heavy-light pseudoscalar mesons are also
given in physical units.

\medskip

To isolate the ground state more efficiently in the correlation
functions that we calculate, we use fuzzed sources and/or sinks
\cite{Lacock:1995qx}. These are extended interpolation operators that have
improved overlap with the ground state.  Of course, operators whose
matrix elements we wish to compute are kept local.

\medskip

Statistical errors are estimated from a bootstrap procedure
\cite{efron}, which involves the creation of 1000 bootstrap samples
from our set of 188 (498) configurations at $\beta=6.2$ ($\beta=6.0$).
Correlators are fitted for each sample by minimizing $\chi^2$.  The
quoted statistical errors are obtained from the central 68\% of the
corresponding bootstrap distribution.

\medskip

To convert our values for decay constants into physical units
we need an estimate of the inverse lattice spacing. The determination
of this quantity is discussed in \sec{sec:chiral}.

\section{Matching and running}
\label{sec:matching}

Because the lattice and the continuum treat ultraviolet modes
differently, the extraction of continuum matrix elements from lattice
calculations requires a matching procedure. Ideally, this matching is
performed nonperturbatively. For our mean-field-improved action,
however, nonperturbative matching coefficients are not available and
we resort to perturbation theory instead. 

\medskip 

The simulation is performed with the fermion action of \eq{eq:sw} with
$\csw=1/u_0^3$ and the standard Wilson gauge action. However, the
fully mean-field improved action would involve normalizing each
occurrence of a link variable in the action by the measured value of
$u_0$.  To recover the results we would have obtained had we used the
latter, we must, in interpreting the results of our simulation, use
rescaled bare couplings $\bar\alpha_s=\alpha_s/u_0^4$ and
$\bar\kappa_\psi=\kappa_\psi u_0$, with
$\alpha_s=g^2_0/(4\pi)$. $\csw$ already has its desired rescaled value
and no additional rescaling is necessary. $\bar\alpha_s$ is actually
only a first guess at an improved expansion parameter and one may try
to optimize this choice \cite{Lepage:1993xa}. This issue will be
elaborated on at the end of the present section and, for the moment,
we will generically denote the coupling by $\alpha_s$.

\medskip

In perturbation theory, the effect of normalizing link variables is
obtained by expanding the factors of $u_0$ in powers of the strong
coupling. At $\ord{\alpha_s}$ we have \cite{Lepage:1993xa},
\beq
u_0\equiv \l\la \frac{1}{3}\tr\upl \r\ra^{\frac{1}{4}}=
1+\frac{\alpha_s}{4\pi} X \mbox{\quad\quad with  } X=-\frac{4\pi^2}{3}
\ .
\eeq
This means that every occurrence of $\kappa_\psi$ in a first order
perturbative expression must be replaced by $\bar\kappa_\psi\
(1-(\alpha_{s}/4\pi) X)$. Because factors of $\csw$
always appear multiplied by at least one power of $\alpha_s$ in
perturbative expressions, $\csw=1-3(\alpha_{s}/4\pi) X$
can be replaced by $\csw=1$ in first order expressions. This
is what we do to determine the central values of our results. However,
in obtaining errors, we consider the variation induced by
taking $\csw=1/u_0^3$.

\medskip

To extract the $B$ parameters of \eq{eq:bparamdef} from ratios
of three-point and two-point correlation functions on the lattice, we
must match the mean-field improved, EKM-normalized 
lattice axial-vector current to its Euclidean continuum counterpart via
\beq
A_\mu = Z_{A}
(\alpha_s)
A_\mu^\lat (a)
\ ,
\label{eq:bilinmatch}
\eeq
with
\beq
A_\mu^\lat (a)\equiv\frac{1}{a^3}
\sqrt{1+a\bar m_q}\,\sqrt{1+a\bar m_Q}\,\sqrt{2\bar\kappa_q\,2\bar\kappa_Q}
\l(\bar Q\gamma_{\mu}\gamma_{5} q\r)(a)
\ ,
\label{eq:bilindef}
\eeq
where $a\bar m_{q,Q}= (1/\bar\kappa_{q,Q}-1/\bar\kappa_{cr})/2$, with
$\bar\kappa_{cr}$ the mean-field improved version of the critical
hopping parameter, $\kappa_{cr}$, which is determined
nonperturbatively as detailed in \sec{sec:chiral}.  Using the results
of Refs.\ \cite{Gabrielli:1991us,Capitani:1997nj}, we find at one
loop\footnote{In practice, we use
\cite{Capitani:1997nj} where results are given for arbitrary 
$\csw$ and where loop integrals are
calculated to higher numerical accuracy.}:
\bea
Z_{A}(\alpha_s) &=& 
1+\frac{\alpha_s}{4\pi}\l[\frac{4}{3}
(\Delta_{\gamma_{\mu}\gamma_{5}}+\Delta_{\Sigma_1})
-X\r]\nn\\
&=& 1-\frac{\alpha_s}{4\pi}[7.90+0.33\csw-3.00\csw^2] 
\ ,
\label{eq:zbilin}
\eea
where, in the notation of Ref. \cite{Gabrielli:1991us},
$\Delta_{\gamma_{\mu}\gamma_{5}}$ and $\Delta_{\Sigma_1}$ arise from
the one-loop corrections to the vertex $\gamma_{\mu}\gamma_{5}$ and to
the quark wave-function, respectively.  The effect of mean-field
improvement is encoded in the term proportional to $X$. Without
mean-field improvement, {\it i.e.} $X=0$, the coefficient of
$\alpha_s/4\pi$ would be substantially larger: 18.39 instead of 5.23
for $\csw=1$.

\medskip

The matching of the four-quark operator $\cO^{\Delta F=2}_q$
($F$ stands for the flavor of the heavy quark) is complicated by the
fact that Wilson-type fermions break chiral symmetry explicitly,
inducing mixing amongst four-quark operators of different chirality.
The following five operators form a complete basis for this mixing on
the lattice in the parity-conserving sector:
\bea
\cO_{1,2}^\lat &=& \gamma_\mu\times\gamma_\mu \pm 
\gamma_\mu\gamma_5\times\gamma_\mu
\gamma_5\ ,\nn\\
\cO_{3,4}^\lat &=& I\times I \pm \gamma_5\times\gamma_5\ ,
\label{eq:opbasis}\\
\cO_5^\lat &=& \sigma_{\mu\nu}\times\sigma_{\mu\nu}\ ,\nn
\eea
where a sum over Lorentz indices is implicit and where
$\Gamma\times\Gamma$ stands for 
$(1/a^{6})$ $\l(1+a\bar m_q\r)$ 
$\l(1+a\bar m_Q\r)$
$\l(2\bar\kappa_q\,2\bar\kappa_Q\r)$
$(\bar Q\Gamma q)(\bar Q\Gamma q)$.  The
parity even component of $\cO^{\Delta F=2}_q$ corresponds to 
the continuum equivalent of $\cO_1^\lat$ so that, at one loop:
\beq
\cO^{\Delta F=2}_q(\mu) \longrightarrow 
Z_{11}(\alpha_s,a\mu)\l(\cO_1^\lat(a)+
\sum_{i=2}^5 Z_{1i}(\alpha_s) \cO_i^\lat(a)\r)\ .
\label{eq:4qmatching}
\eeq
$Z_{11}$ has a logarithmic dependence on $a\mu$, where $\mu$ is the
scale at which the continuum operator is renormalized, while $Z_{1i}$,
$i=2,\cdots,5$, remains finite as $a$ vanishes.  Mean-field improving the
dimensional reduction (DRED) results of Refs.
\cite{Gabrielli:1991us,Frezzotti:1992pe,Capitani:1997nj} and using the
matching between DRED and NDR-$\msbar$ given in Ref. 
\cite{Gupta:1997yt}, we
find that one-loop matching to the NDR-$\msbar$ scheme is given by:
\bea
Z_{11}(\alpha_s,a\mu) &=& 
1+ \frac{\alpha_s}{4\pi}\bigg [ -4\,\ln(a\mu) + \frac{1}{3}
\big\{5(\Delta_{\gamma_\mu}+\Delta_{\gamma_\mu\gamma_5})\nn\\
& &-
(\Delta_{I}+\Delta_{\gamma_5})+8\Delta_{\Sigma_1}
\big\}-2-2X\bigg ]\label{eq:4qzs} \\
&=& 1-\frac{\alpha_s}{4\pi}[4\,\ln(a\mu)+24.52-9.33\csw-4.88\csw^2]\,\ ,\nn\\
Z_{12}(\alpha_s)&=& 
-\frac{11}{12}\frac{\alpha_s}{4\pi}\l[\Delta_{I}-\Delta_{\gamma_5}
\r]=\,-\frac{\alpha_s}{4\pi}[8.84-9.15\csw+3.13\csw^2]\ , 
\nn
\\
Z_{13}(\alpha_s) &=& -\frac{4}{11} Z_{12}(\alpha_s)\,,\,
Z_{14}(\alpha_s) = \frac{2}{11} Z_{12}(\alpha_s)\,, \,
Z_{15}(\alpha_s) = \frac{2}{11} Z_{12}(\alpha_s)
\ ,\nn
\eea
where the $\Delta_{\Gamma}$, $\Gamma=\gamma_\mu$,
$\gamma_\mu\gamma_5$, $I$ and $\gamma_5$ arise from the one-loop
corrections to the bilinear vertices associated with
$\Gamma$.~\footnote{Here again, we use the more precise and general
$\csw$ results of Ref. \cite{Capitani:1997nj} for
$\Delta_{\Gamma}$.}

\medskip

For the numerical evaluation of the renormalization constants, we choose
to work with the $\msbar$ coupling, $\alpha_{\msbartiny}$, obtained via
\beq
\alpha_{\msbartiny}(3.41/a) = \alpha_V({\mathrm{e}}^{5/6}3.41/a)
\l (1+2\frac{\alpha_{V}}{\pi}\r ) \ ,
\label{eq:alphamsbar}\eeq
where $\alpha_V$ is the coupling defined from the heavy-quark potential.
The latter is obtained from our simulations by solving \cite{Lepage:1993xa}
\beq
\mbox{ln}(u_0) = \frac{\pi}{3}\alpha_V(3.41/a)\l(1-1.185\alpha_V\r)
\label{eq:alphav}
\ .\eeq
Values of $\alpha_{\msbartiny}$ for scales different from $3.41/a$
are obtained by solving the two-loop running equation numerically, with
$n_f=0$. Both $\alpha_{\msbartiny}$ and $\alpha_V$ have been shown to
lead to expansions that are much more convergent than those in terms
of the bare lattice coupling \cite{Lepage:1993xa}. For completeness, we give
values of $\alpha_{\msbartiny}(\mu)$ for a collection of $\mu$ in
\tab{tab:alphav}.  

\medskip

Having chosen the coupling, we must fix the scale, $q^*$, at which
it is evaluated. We take $2/a$ as a central value but allow $q^*$ to
vary from $1/a$ to $\pi/a$ to estimate the uncertainty associated with
this choice.  This range for $q^*$ covers typical, ultraviolet lattice
scales. 

\medskip

As it is convenient for the heavy-quark extrapolations and does
not generate large logarithms, we match the lattice results for the
$B$ parameters at $\mu=M_{P_l}$, where $M_{P_l}$ is the mass of the
heavy-light meson composed of a heavy antiquark $\bar Q$ and a
massless quark, $l=d$ or $u$. As we will see in \sec{sec:hq}, the
heavy-quark extrapolation yields $B_{B_q}(M_{B_d})$.  Values for the
renormalization-group-invariant and scheme-independent $B$ parameters
are then simply obtained from \eqs{eq:bnlodef}{eq:cbdef}: $\hat
B_{B_q}^\nlo=C_B(M_{B_d}) B_{B_q}(M_{B_d})$.  Values for
$B_{B_q}(\mu)$, with $\mu\ne M_{B_d}$, can also be obtained
straightforwardly through: $B_{B_q}(\mu)=\l[C_B(M_{B_d})/C_B(\mu)\r]$
$B_{B_q}(M_{B_d})$, with or without an expansion in $\alpha_s$ of the
term proportional to $J_5$ in \eq{eq:cbdef}

\medskip

Since we match the matrix elements and $B$ parameters defined in
\eq{eq:bparamdef} at one loop, we may choose to expand combinations of
renormalization constants such as $Z_{11}/Z^{2}_{A}$ or
$Z_{11}Z_{1i}$, $i=2,\cdots,5$, to order $\alpha_s$. In the present
paper, all central values are obtained without
expansion. Nevertheless, we have checked that expanding these
combinations makes negligible differences in the final results. This is
because the one-loop corrections to the renormalization constants are
small, especially after mean-field improvement.

\section{Correlation functions}
\label{sec:corfunc}

To determine $\Delta F=2$ matrix elements and their $B$ parameters, as
well as decay constants, we compute the following two- and three-point
functions:
\bea 
C_{PP}^{FF}(t_x) &=& \sum_{\vec{x}} \la P_F^{\dagger}(x)P_F(0)\ra\ 
,\nn\\ 
C_{AP}^{LF}(t_x) &=& \sum_{\vec{x}} \la
P_F^\dagger(x)A_0^\lat(0)\ra\ ,
\label{eq:2and3pt}\\
C_{\cO}^{FLF}(t_x,t_y) &=& \sum_{\vec{x},\vec{y}}
\la P^\dagger_F(y)\cO(0)P^\dagger_F(x)\ra\ ,\nn
\eea
as well as $C_{PA}^{FL}(t_x)$ and $C_{PP}^{FL}(t_x)$. Here, the
subscripts $F$ on operators indicate that they are fuzzed
\cite{Lacock:1995qx}, while operators with no $F$ are local. For the
correlation functions, the superscripts $F$ and $L$ indicate which of
their operators are fuzzed ($F$) or local ($L$).  In \eq{eq:2and3pt},
$A_0^\lat$ is the time component of the EKM-normalized axial-vector
current defined in \eq{eq:bilindef}, $P_F$ is a fuzzed \cite{Lacock:1995qx}
version of the pseudoscalar density $\bar Q\gamma_5 q$, and
$\cO$ stands for any of the four-fermion operators in
\eqs{eq:opbasis}{eq:4qmatching}.  In the present section, we
set the lattice spacing $a=1$ and omit light-quark indices on
heavy-light quantities, for notational simplicity.

\medskip

\begin{figure}
\centerline{\epsfxsize=0.5\textwidth\epsffile{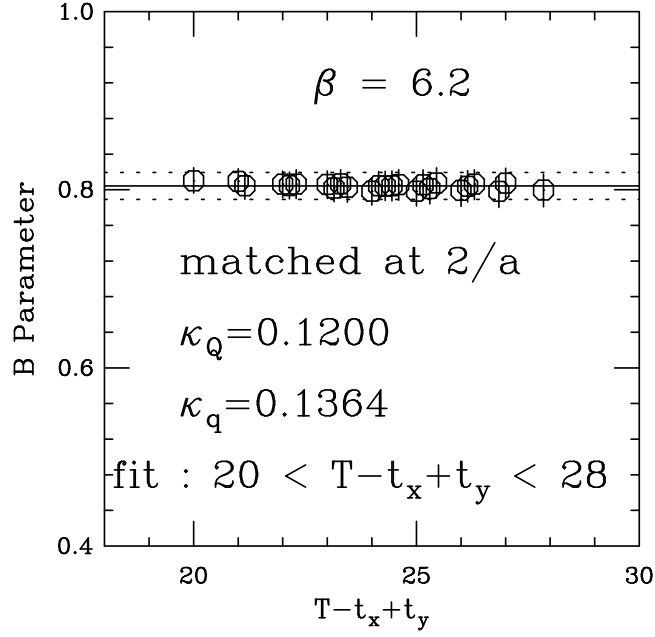}}
\caption{\label{fig:plat} Plateau for the $B$-parameter of the
operator $\cO^{\Delta F=2}_q$ renormalized at $\mu=2/a$ in the
NDR-$\msbar$ scheme, for $\beta=6.2$ (i.e. \eq{eq:corrratsB} times
$3/(8Z_A^2)$).  The correlation function is obtained for $10
\le t_{y} \le 14$ and $10 \le T-t_{x} \le 14$.  Points with the same
$T-t_{x}+t_{y}$ are shifted for clarity.}
\end{figure}

At large Euclidean time separations, the three-point correlation function
$C_{\cO}^{FLF}(t_{x},t_{y})$ has the asymptotic
behavior~\footnote{On our periodic lattices this corresponds to
  $at_{y}$ and $a(T-t_{x})$ much greater than 1 but small enough so
  that the desired time ordering dominates. Here, $aT=48$ is the time
  extent of our lattices. We consider times $t_x$
  and $t_y$ such that $T/2<t_x<T$ and $0<t_y<T/2$.}
\beq 
C_{\cO}^{FLF}(t_{x},t_{y}) \to \frac{1}{4M_P^2} \la 0 |
P^{\dagger}_F | \bar{P} \ra \la \bar{P} | \cO | P \ra \la P
| P^{\dagger}_F | 0\ra \exp \l [ -M_{P}(T-t_{x}+t_{y})\r ] \ 
, \label{eq:asymp3pt}
\eeq
where $M_{P}$ is the mass of the pseudoscalar $P$.  Therefore, fits to
the ratios of correlation functions:
\bea 
\frac{C_{\cO^{\Delta
      F=2}}^{FLF}(t_x,t_y)}{C_{AP}^{LF}(t_x) C_{AP}^{LF}(t_y)} &
\longrightarrow & \frac{8}{3}Z_{A}^2 B_P\ , \label{eq:corrratsB}\\ 
\frac{C_{\cO^{\Delta F=2}_q}^{FLF}(t_x,t_y)}{C_{PP}^{FF}(t_x)
  C_{PP}^{FF}(t_y)} & \longrightarrow & \frac{\la\bar
  P|\cO^{\Delta F=2}_q| P\ra}{\la
  0|P_F|P\ra \la
  \bar P|P_F|0\ra}\label{eq:corrratsml}\ ,
\eea
where $B_P$ is the $B$ parameter corresponding to the heavy-light
meson $P$, yield the desired quantities up to renormalization
constants that we determine perturbatively (see \sec{sec:matching}) and
factors of $M_P$ and of the type, $\la 0| P_F | P\ra$, that we
determine from fits to the two-point functions
$C_{PP}^{FL}(t_x)$\footnote{$C_{PP}^{FL}$ gives a particularly good
signal for $M_P$.} and $C_{PP}^{FF}(t_x)$, respectively.  Note that in
\eqs{eq:asymp3pt}{eq:corrratsml}, all pseudoscalar meson states have
vanishing three momentum. An example of a plateau for the
ratio of \eq{eq:corrratsB} is shown in \fig{fig:plat}.

\medskip

To determine the decay constants, we consider
\beq
\frac{C_{PA}^{FL}(t_x)}{C_{PP}^{FF}(t_x)} \longrightarrow 
\frac{M_P\, f_P}{Z_A\la 0|P_F^\dagger|\bar P\ra}{\mathrm{tanh}}
\l[M_P(T/2-t_x)\r]\ .
\label{eq:corrratdc}
\eeq

\medskip

In order to investigate the dependence of the matrix elements and
$B$ parameters on heavy- and light-quark mass, we need the
heavy-light and the light-light pseudoscalar meson masses, and the
light-light decay constants, which we use to set the scale. As already
stated, we obtain the heavy-light pseudoscalar meson mass, $M_{P}$,
from a fit to the large time behavior of $C_{PP}^{FL}(t_x)$.
Similarly, we get the light-light pseudoscalar meson mass, $m_p$, from
a study of the two-point function $C_{pp}^{FL}(t_x)$ where $p=\bar
q_1\gamma_5 q_2$ is the pseudoscalar bilinear made from the two light
quarks $q_1$ and $q_2$. Finally, the light-light pseudoscalar decay
constant is obtained from a fit to the ratio
$C_{pa}^{FL}(t_x)/C_{pp}^{FF}(t_x)$, where $a_0$ is the
EKM-normalized version of $\bar q_1\gamma_0 \gamma_5 q_2$, much in
the same way the heavy-light decay constant is obtained.

\medskip

The time ranges over which the various correlation functions and
ratios are fitted are given in \tab{tab:calcandfittimes}.

\section{Light-quark-mass extrapolations and determination of the 
lattice spacing}
\label{sec:chiral}

\begin{figure}
\centerline{\begin{tabular}{cc}
\epsfxsize=0.48\textwidth\epsffile{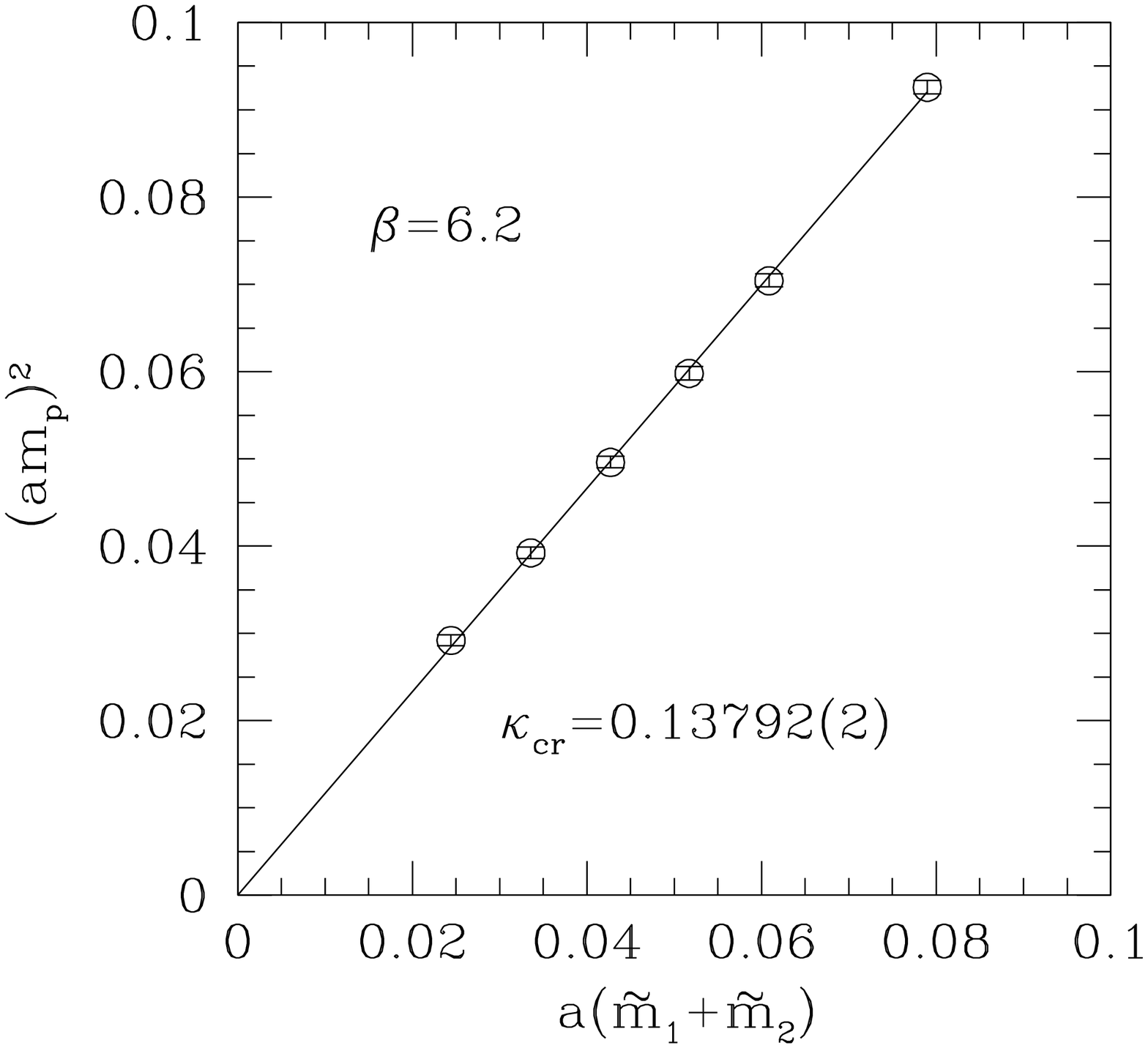}&
\epsfxsize=0.48\textwidth\epsffile{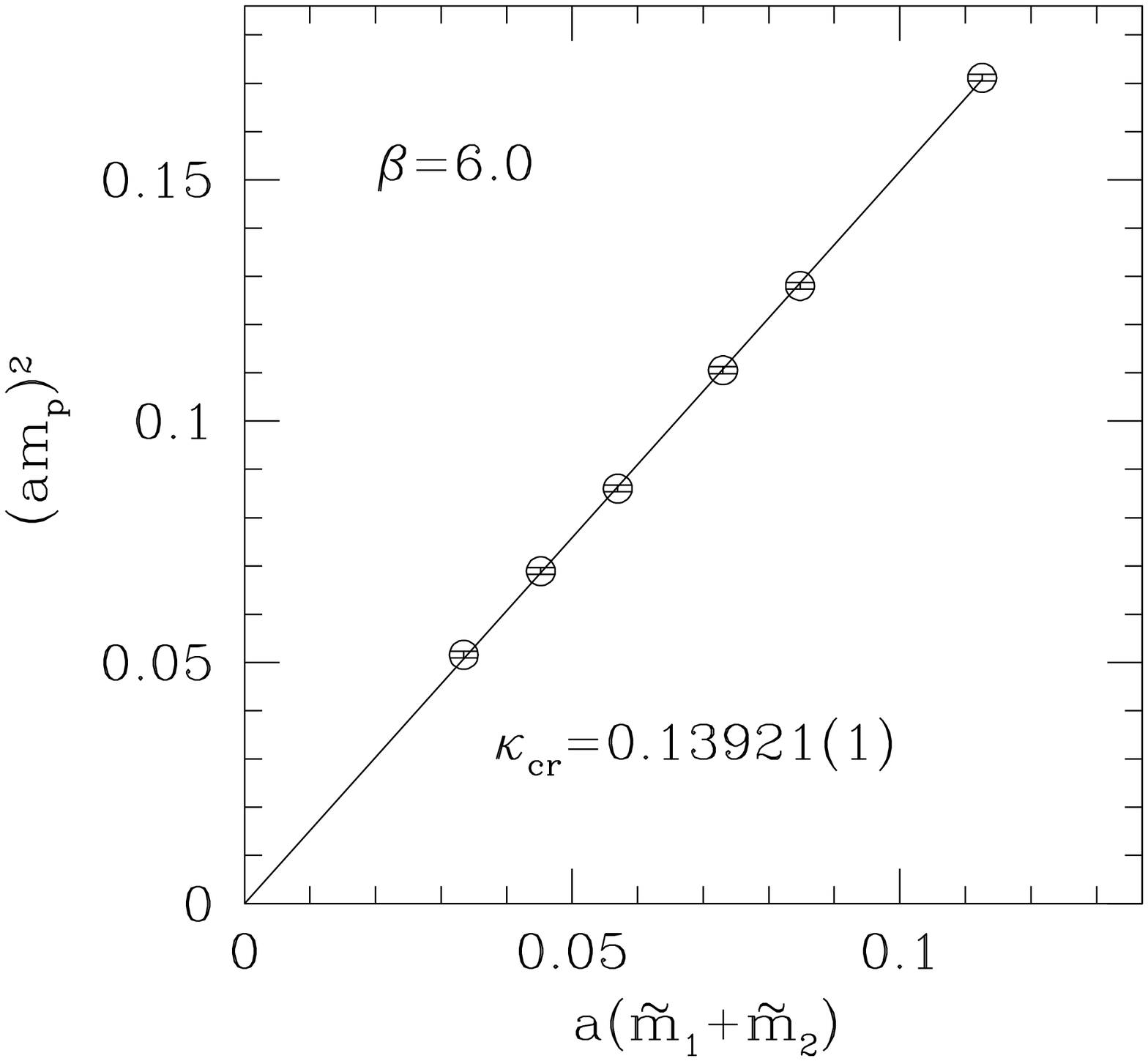}
\end{tabular}}
\vspace{-0.8cm}
\caption{\label{fig:kappac} Fits of the squared, light-light 
pseudoscalar meson masses versus light-quark mass to the PCAC relation
of \eq{eq:pcac}, at $\beta=6.2$ and 6.0.}
\end{figure}

Results for physical $B_d$ and $B_s$ mesons require investigations of
the dependence of the lattice measurements on light- and
heavy-quark masses.  We begin by the light-quark-mass extrapolations
and interpolations.  To obtain $\kappa_{cr}$, the critical value of the
quark hopping parameter, we study the behavior of the light-light
pseudoscalar meson mass, $m_p$, as a function of $\kappa_1$ and $\kappa_2$,
the hopping parameters of the light quarks which compose it. We assume
that it obeys the partially conserved axial-current (PCAC) relation
\beq
\l(am_p\r)^2(\kappa_1,\kappa_2) =
\beta_m (a\tilde m_1+a\tilde m_2)
\ ,
\label{eq:pcac}
\eeq
where $\tilde m_i$ is the $O(a)$-improved quark mass, given by
\beq
\tilde m_i=m_i(1+b_m am_i)
\ ,
\label{eq:mimprov}
\eeq
with $am_i=1/(2\kappa_i)-1/(2\kappa_{cr})$ \cite{Luscher:1996sc}. At tree
level, which is sufficient with our mean-field-improved action,
$b_m=-1/2$.  $\kappa_{cr}$ is then obtained by fitting the function of
\eq{eq:pcac} to our results for $(am_p)^2$, with $\beta_m$ and
$\kappa_{cr}$ as fit parameters.  $\kappa_{1}$ and $\kappa_{2}$ are
taken amongst the values in the row labeled $\kappa_{q}$ in
\tab{tab:simparam}.  Addition of a term quadratic in quark mass in
\eq{eq:pcac} makes very little difference to the central value for
$\kappa_{cr}$ and this variation has a negligible effect on the
light-quark-mass extrapolations of the matrix elements of
interest. The linear extrapolation of \eq{eq:pcac} is shown in
\fig{fig:kappac}.

\medskip

For $m_p$ and all the quantities we study here, in addition to
higher-order polynomial corrections in light-quark mass, chiral
perturbation theory also predicts the presence of chiral logarithms.
These logarithms, however, are difficult to isolate numerically and
are modified by the quenched approximation
\cite{Sharpe:1992ft,Bernard:1992mk,Sharpe:1996qp}. Thus, we perform
polynomial interpolations and extrapolations from our intermediate
values of light-quark mass, only considering in most cases the leading,
non-trivial dependence on light-quark mass.

\medskip

\begin{figure}
\centerline{\begin{tabular}{cc}
\epsfxsize=0.48\textwidth\epsffile{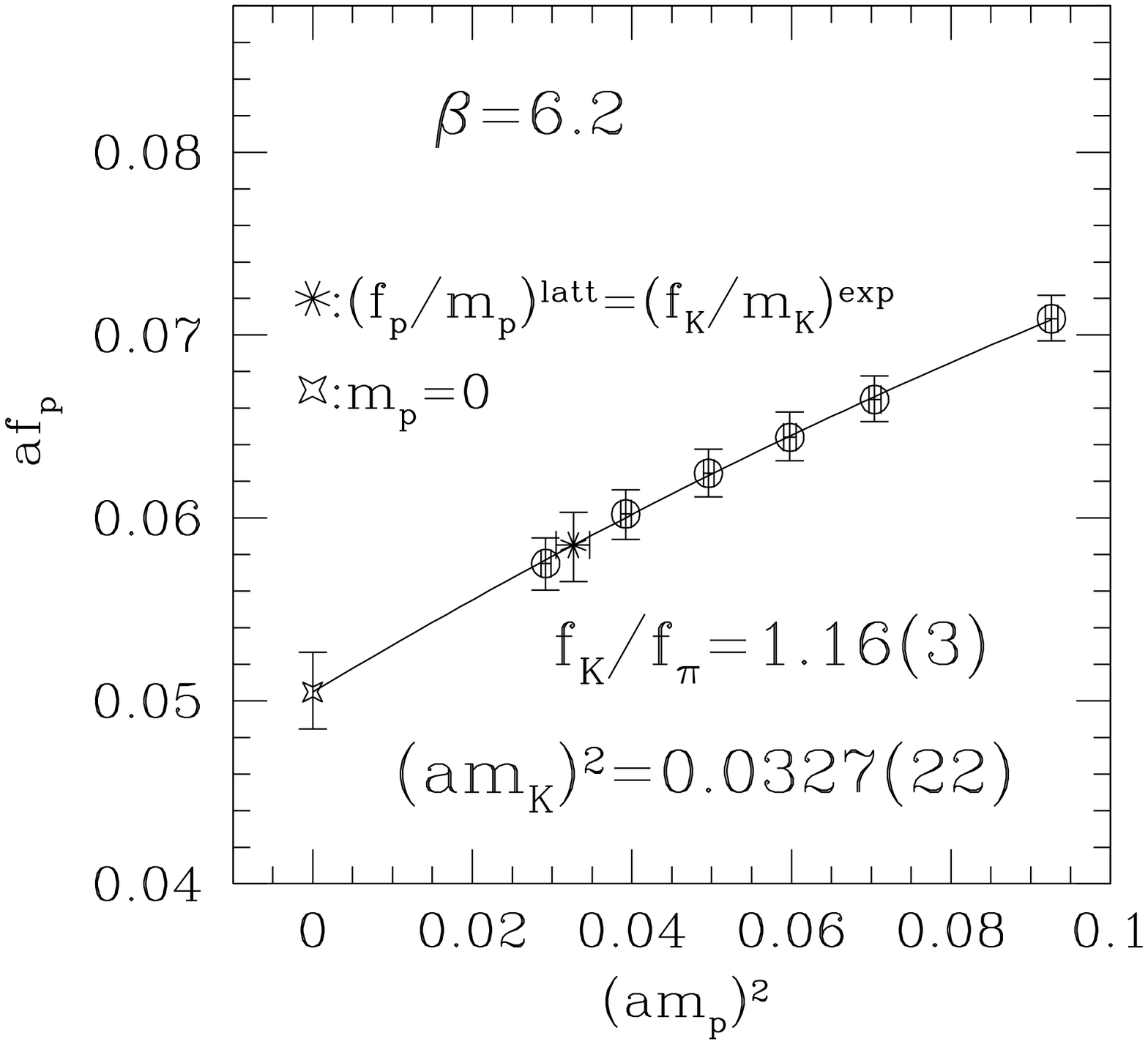}&
\epsfxsize=0.48\textwidth\epsffile{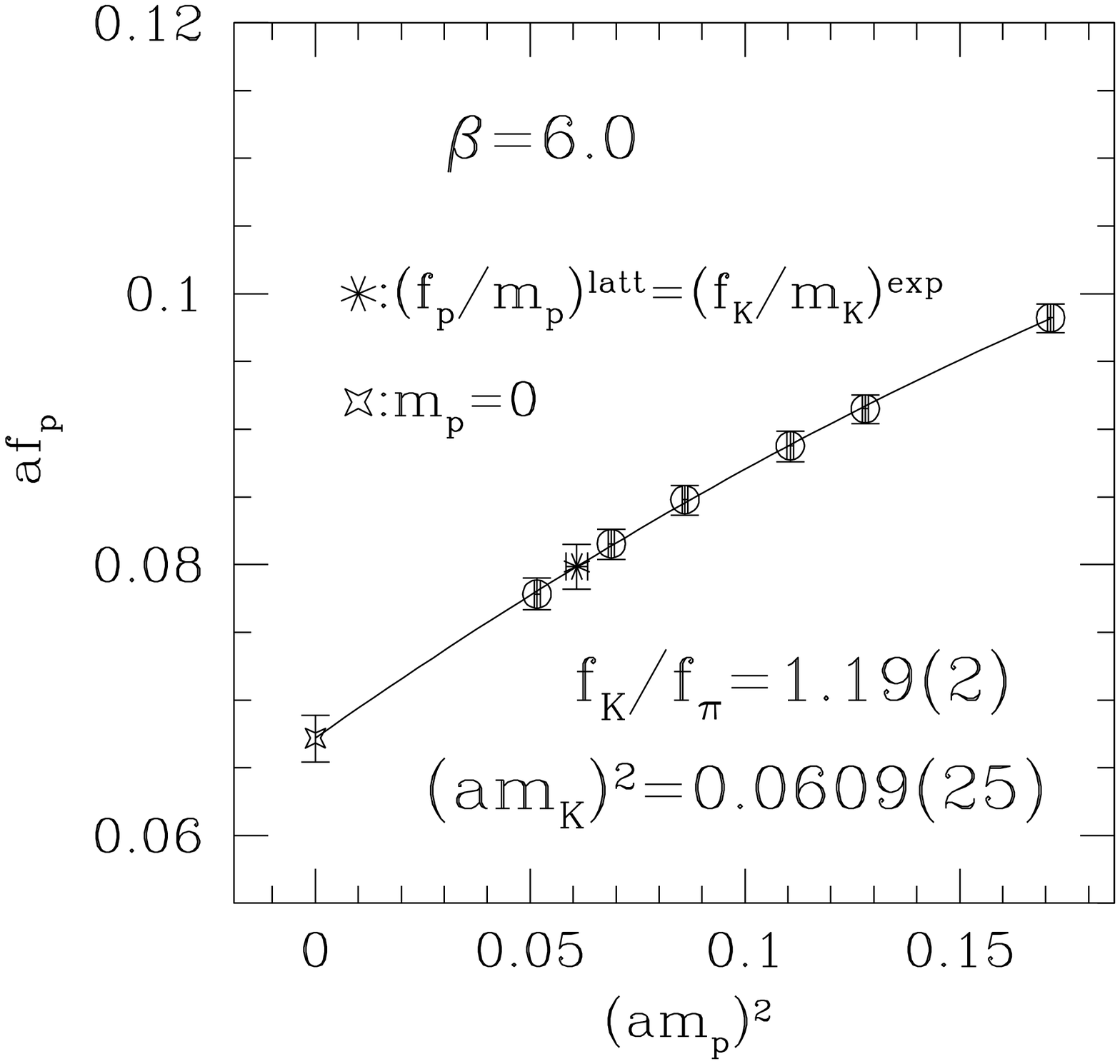}
\end{tabular}}
\vspace{-0.8cm} \caption{\label{fig:fkscale} Interpolation of
$af_p$, according to \eq{eq:fkscale}, which is used to determine the
scale $a^{-1}$, at $\beta=6.2$ and 6.0, from $f_K$ and $m_K$, as described
in the text.}
\end{figure}

Next we determine the lattice scale $a^{-1}$ and the bare
strange-quark mass, $a\tilde m_s$, in lattice units.  We obtain both
quantities simultaneously from the kaon's decay constant, $f_K$, and
its mass, $m_K$.  As long as the mass dependence of $f_p$, the decay
constant of a light-light pseudoscalar meson, can be described by a
function of $m_p^2$ only,
\eq{eq:pcac} implies that both quantities depend only on the sum of
quark masses and that only an interpolation in these masses, not an
extrapolation, is needed to obtain $af_K$. The use of $f_K$,
instead of, for instance, $m_{K^*}$, is prompted by the fact that the
only quantities we report on here which depend strongly (i.e. not
logarithmically) on $a^{-1}$ are the heavy-light decay constants. Our
procedure has the added benefit that it is also applicable in
unquenched simulations~\footnote{The $K^*$ is not a stable particle
  once light-quark loops are allowed.}. Thus, we fit our results for
$af_p/Z_A$ to:
\beq
Z_A\l(\frac{af_p}{Z_A}\r) = 
\alpha_f+\beta_f(am_p)^2+\gamma_f(am_p)^4+\cdots
\ .
\label{eq:fkscale}
\eeq
We find that this parametrization describes our results well.
Assuming that the lattice spacing is fixed with $f_K$ (or equivalently
$m_K$), we solve \eq{eq:fkscale} for $af_p$ (or $am_p$), at the point
specified by the physical ratio $m_K/f_K$, with $f_K=159.8\mev$ and
$m_K=493.7\mev$. $a\tilde m_s$ is then obtained from the resulting
$am_K$, using our earlier fit to \eq{eq:pcac}, and the lattice spacing
from the resulting $af_K$ (or $am_K$). The fits at our two values of the
coupling are shown in \fig{fig:fkscale}.  Because of the slight
curvature, we favor the quadratic fits. These fits also give a value
of $f_K/f_\pi$ which is closer to the experimental result of 1.22:
$1.16(3)$ ($1.19(2)$) for quadratic fits instead of $1.131(10)$
($1.147(7)$) for linear fits, at $\beta=6.2$ (6.0). In any case,
linear and quadratic fits give nearly identical results for the scale
and $a\tilde m_s$. The values of $Z_A$ used are those obtained from
\eq{eq:zbilin} with $\alpha_s=\alpha_{\msbartiny}(2/a)$. Systematic
uncertainties in the determination of $a^{-1}$ and $a\tilde m_s$ will be
addressed when we discuss the uncertainties on our $B$-physics results
in \sec{sec:results}.  A summary of the results for $a^{-1}$, $\kappa_{cr}$
and $a\tilde m_s$ used below is given in \tab{tab:spectrum}.

\medskip

\begin{figure}
\centerline{\begin{tabular}{cc}
\epsfxsize=0.48\textwidth\epsffile{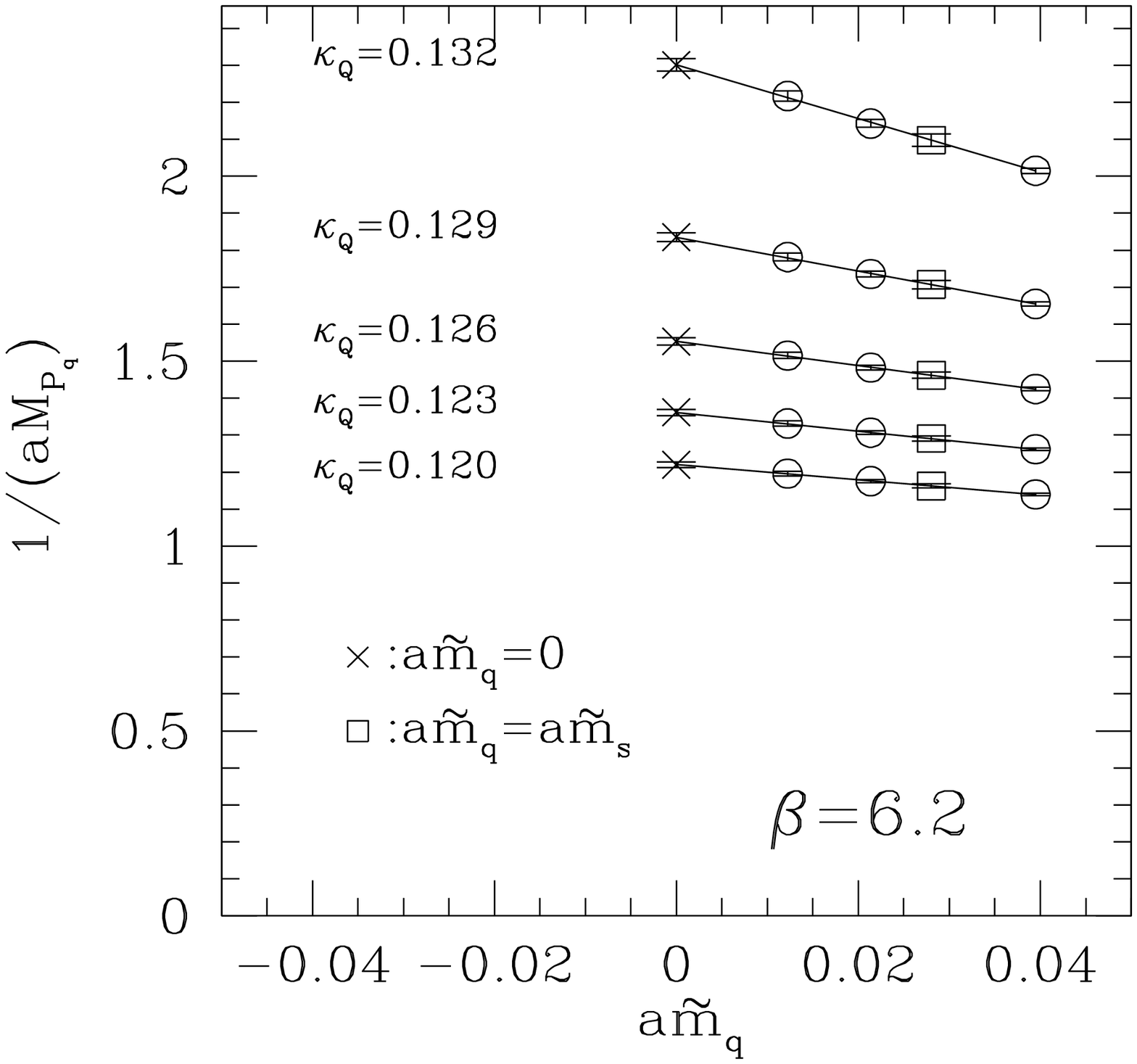}&
\epsfxsize=0.48\textwidth\epsffile{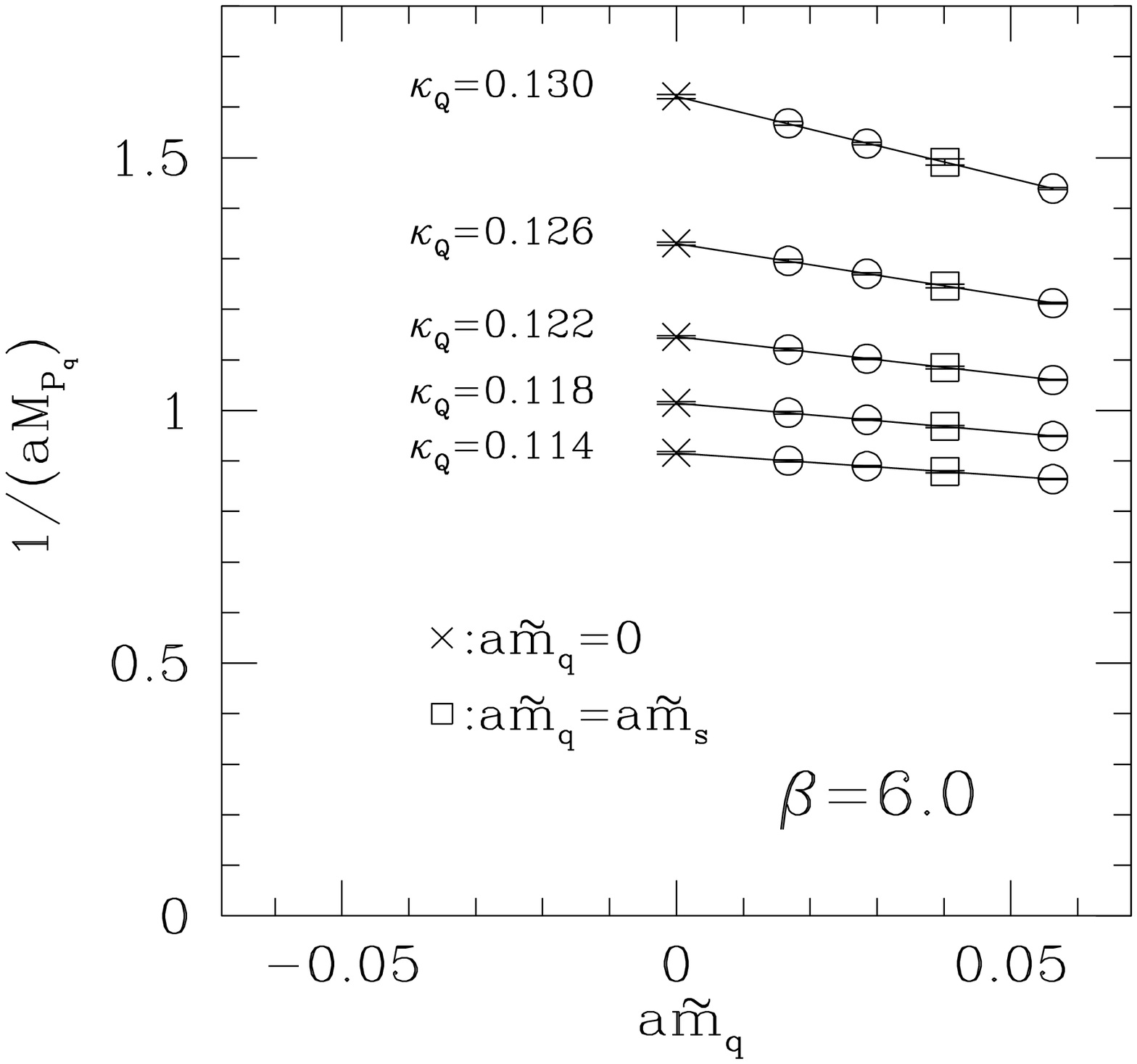}
\end{tabular}}
\vspace{-0.8cm}
\caption{\label{fig:mpvslqm}
Light-quark-mass dependence of the heavy-light pseudoscalar meson mass,
$M_{P_q}$, and extrapolation and interpolation to vanishing quark mass and
strange-quark mass at $\beta=6.2$ and 6.0.}
\end{figure}

Having determined the strange quark mass and critical hopping
parameter, we interpolate and extrapolate our various heavy-light
matrix elements in light-quark mass to these values. We assume that
the up and the down quarks are massless. This is an excellent
approximation for the quantities we study. As mentioned above, we
perform polynomial interpolations and extrapolations from our values
of light-quark mass.  Thus, we fit all quantities of interest, $Y$, in
lattice units, to the functional form:
\beq
Y(\kappa_Q,\kappa_q) = \alpha_Y+\beta_Y a\tilde m_q
+\cdots
\ .
\label{eq:lqmdep}
\eeq
In \figsss{fig:mpvslqm}{fig:Bvslqm}{fig:matvslqm}{fig:fpvslqm} we
exhibit the light-quark-mass dependence and the corresponding fits to
\eq{eq:lqmdep} for the quantities $1/(aM_{P_q})$, $B_{P_q}$,
$a^4\la\bar P_q|O^{\Delta F=2}_q| P_q\ra$ and $af_{P_q}/Z_A$, for all
values of heavy-quark mass at $\beta=6.0$ and $6.2$.  For clarity of
presentation, the $B$ parameters and $a^4\la\bar P_q|O^{\Delta F=2}_q|
P_q\ra$ are renormalized at a common scale of 5 GeV~\footnote{These
values of $B_{P_q}$ and $a^4\la\bar P_q|O^{\Delta F=2}_q| P_q\ra$ are
obtained by matching the lattice results onto the NDR-$\msbar$ scheme
at the scale $2/a$, then running to 5 GeV in the $\msbar$ scheme at the
two-loop level with $n_{f}=0$ and the coupling constant described in
Section \ref{sec:matching}.}.  In all cases, the light-quark-mass
dependence is mild and, to good accuracy, linear.  However, we do
observe that the matrix elements of $O^{\Delta F=2}_q$ have a stronger
dependence on light-quark mass than the other quantities.

\section{Heavy-quark-mass extrapolations}
\label{sec:hq}

\begin{figure}
\centerline{\epsfxsize=0.5\textwidth\epsffile{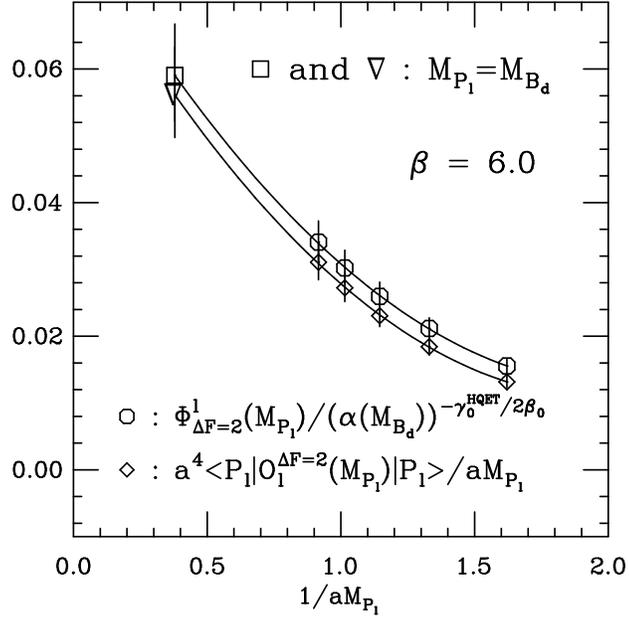}}
\caption{\label{fig:hqetlogs} Influence of leading logarithms on
heavy-quark-mass scaling: behavior of 
$a^{4}\la\bar P_l|\cO^{\Delta F =
2}_l(M_{P_l})|P_l\ra/aM_{P_l}$ (i.e. logarithms omitted) 
and of $\Phi_{\Delta F =
2}^l(M_{P_l})/\alpha_s(M_{B_d})^{4/11}$ 
(i.e. leading logarithms included) versus $1/(aM_{P_l})$.}
\end{figure}

The second extrapolation we have to perform is in heavy-quark mass.
We use heavy quark effective theory (HQET) as a guide, with $M_{P_l}$
as a measure of the heavy-quark mass everywhere except for kinematical
dependencies, where the appropriate meson mass is used. As before, 
$l$ stands for either a $d$ or a $u$ quark. Other choices
for the heavy-quark scale, such as $M_{P_s}$, for instance, make very
little difference to the final results. We match our QCD results
onto HQET at the heavy-quark scale $M_{P_l}$, and 
cancel the leading logarithmic dependence on
$M_{P_l}$ by including terms of the form
$[\alpha_s(M_{P_l})]^{-\gamma^{\hqettiny}_0/2\beta_{0}}$,
where $\gamma^{\hqettiny}_0$ are the relevant, one-loop
anomalous dimensions in HQET, and $\beta_{0}$ is the one-loop
$\beta$-function coefficient. $\gamma^{\hqettiny}_0=-4$ and
$-8$ for the decay constant and four-quark matrix element,
respectively \cite{RussianBB, PolitzerWise}. Thus, we define
\beq
 \Phi_f^q(M_{P_l}) \equiv
\frac{af_{P_q}}{Z_A}\sqrt{aM_{P_q}}
\times[\alpha_s(M_{P_l})]^{2/\beta_0}
\ ,
\label{eq:phifdef}
\eeq
for the decay constants,  
\beq
 \Phi_B^q(M_{P_l}) \equiv
B_{P_q}(M_{P_l})\times[\alpha_s(M_{P_l})]^{0/\beta_0}
\ ,
\label{eq:phiBdef}
\eeq
for the $B$ parameters, and
\beq
\Phi_{\Delta F=2}^q(M_{P_l})  \equiv  
a^4\la\bar P_q|\cO^{\Delta F=2}_q(M_{P_l})|P_q\ra\frac{1}{aM_{P_q}}
\times[\alpha_s(M_{P_l})]^{4/\beta_0}
\ ,
\label{eq:phidf2def}
\eeq
for the $\Delta F=2$ matrix element. Similar scaling functions are
defined for $SU(3)$-breaking ratios, in which the leading logarithmic
dependence on heavy-quark mass cancels.  In
\eqss{eq:phifdef}{eq:phiBdef}{eq:phidf2def}, we evaluate $\alpha_s(M)$
through $2\pi/(\beta_0\mbox{ log}(M/\Lambda_{\mathrm{QCD}}))$ with
$\Lambda_{\mathrm{QCD}}=100\mbox{ MeV}$ and $\beta_0=11{-}(2/3)n_f$ with
$n_f=0$, since we are working in the quenched approximation. This
one-loop coupling approximates the lattice couplings defined through
\eqs{eq:alphamsbar}{eq:alphav} rather well, for the values of $M$
required here. In fact, final results depend weakly on the value
$\Lambda_{\mathrm{QCD}}$ used in the heavy-quark extrapolations (see
also \fig{fig:hqetlogs} and discussion below).

\medskip

For $X(M_{P_l})=\Phi_{\Delta F=2}^q(M_{P_l})$,
$\Phi_{B}^q(M_{P_l})$, $\Phi_f^q(M_{P_l})$ and the corresponding
scaling functions for $SU(3)$-breaking ratios, we use the
HQET-inspired relation
\beq X(M_{P_l}) =
A_X\l[1+B_X\l(\frac{1}{aM_{P_l}}\r)+C_X\l(\frac{1}{aM_{P_l}}\r)^2+\cdots\r]
\label{eq:hqscal}
\eeq
to investigate the heavy-quark-mass scaling of these quantities.  The
leading logarithms make little difference in the extrapolation, as
shown in \fig{fig:hqetlogs} where we plot the extrapolation of the
$\Delta F=2$ matrix element, which has the strongest logarithmic
dependence amongst the quantities we study, with and without these
logarithms.

\medskip

\begin{figure}
\centerline{\begin{tabular}{cc}
\epsfxsize=0.48\textwidth\epsffile{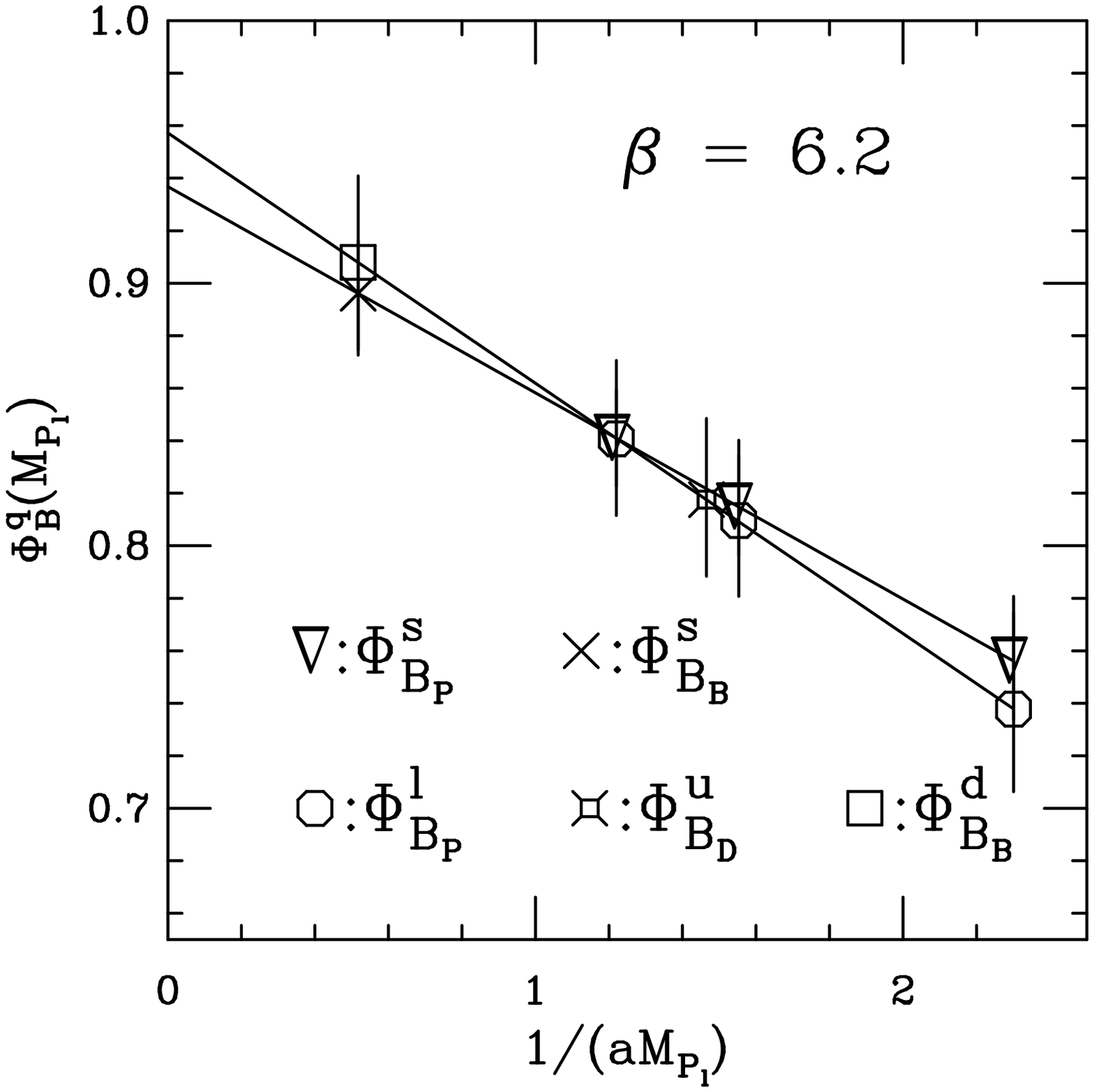}&
\epsfxsize=0.48\textwidth\epsffile{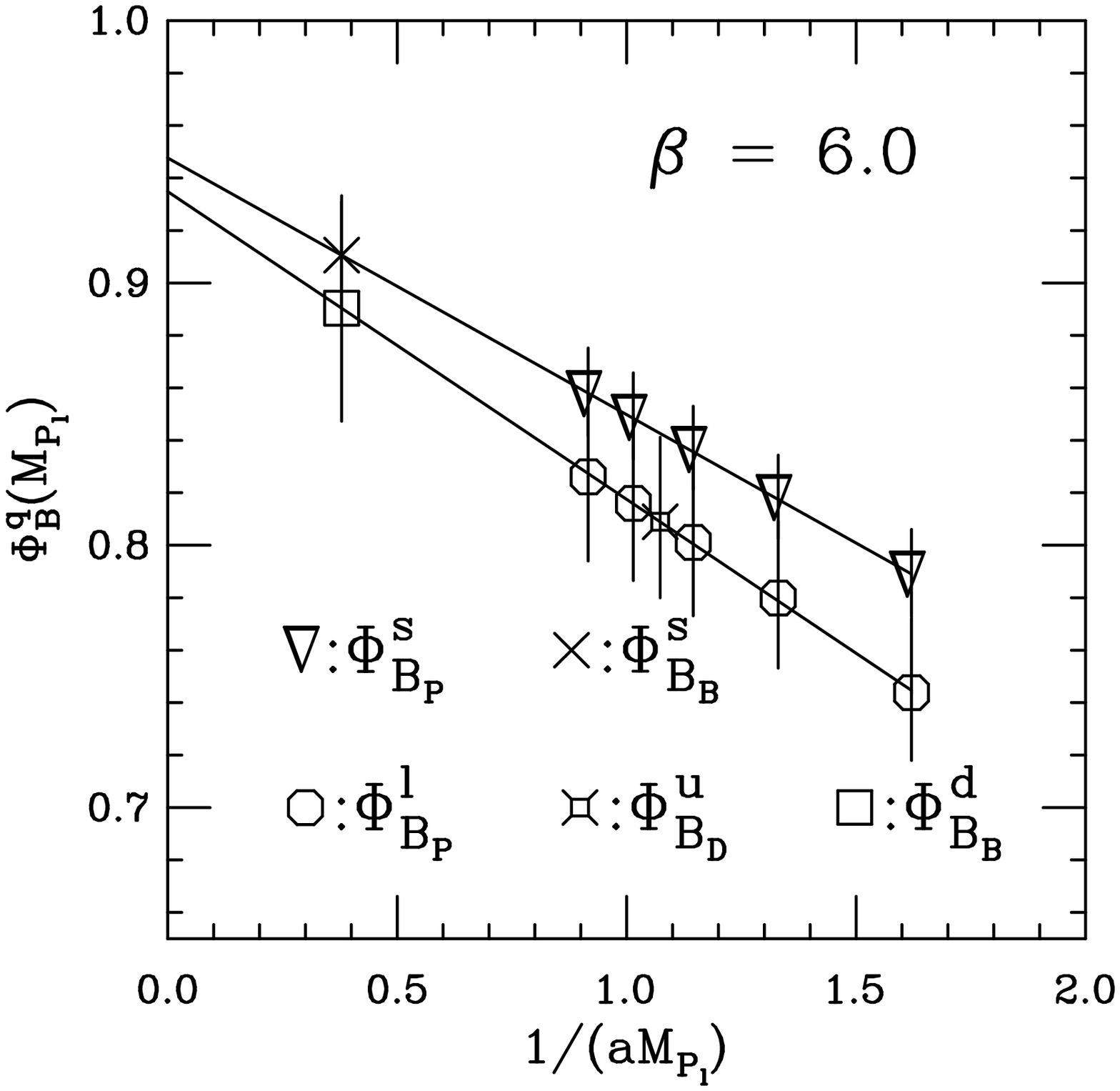}
\end{tabular}}
\caption{\label{fig:Bvshqm}
Lattice results for 
$XM_{P_l}() =\Phi_B^q(M_{P_l})$ versus $1/(aM_{P_l})$ at 
$\beta=6.2$ and 6.0
and for $q=s$ and $l$. The solid lines
are fits to the linear part of the heavy-quark-mass dependence given
in \eq{eq:hqscal}.}
\end{figure}

\begin{figure}
\centerline{\begin{tabular}{cc}
\epsfxsize=0.48\textwidth\epsffile{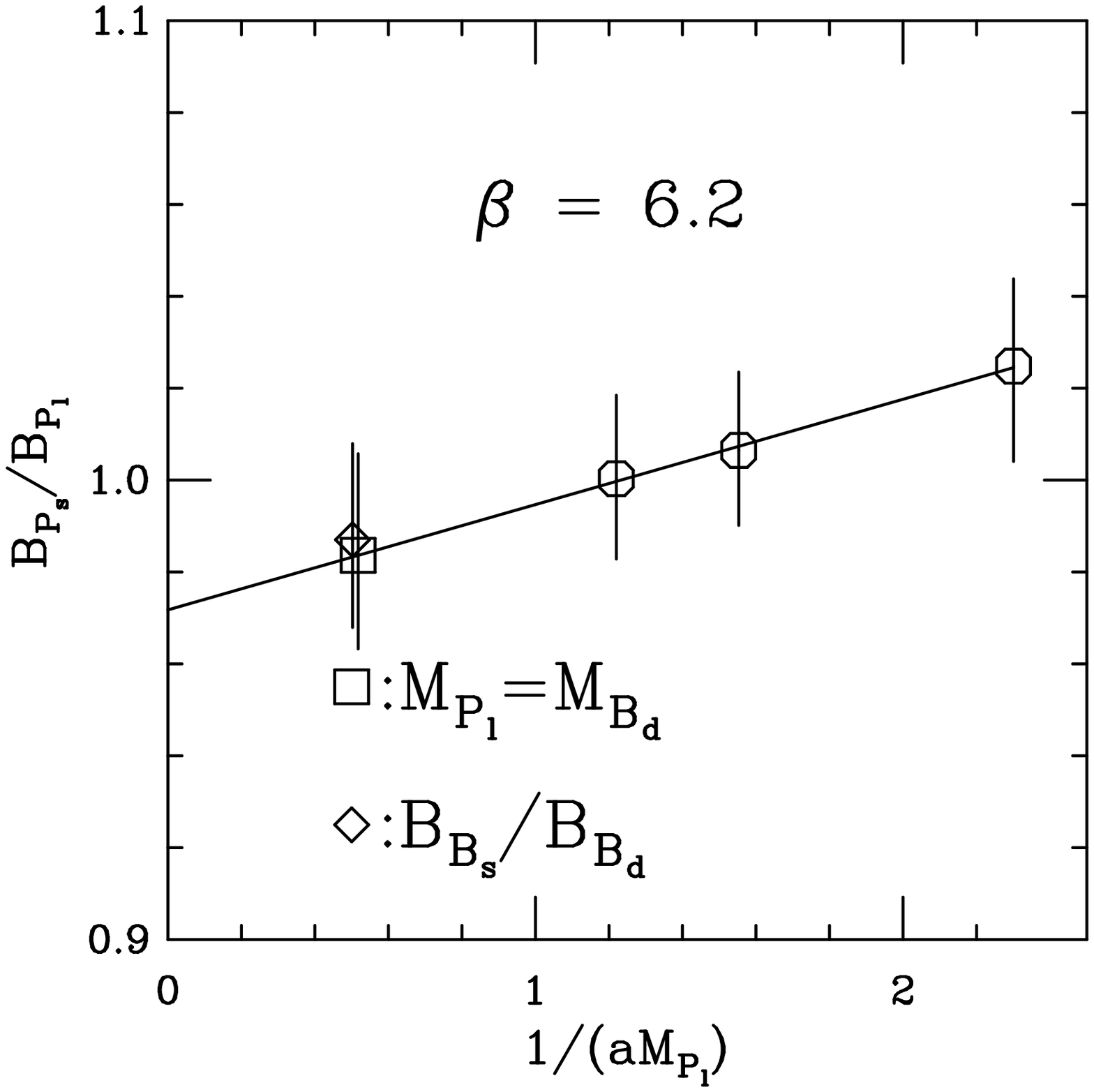}&
\epsfxsize=0.48\textwidth\epsffile{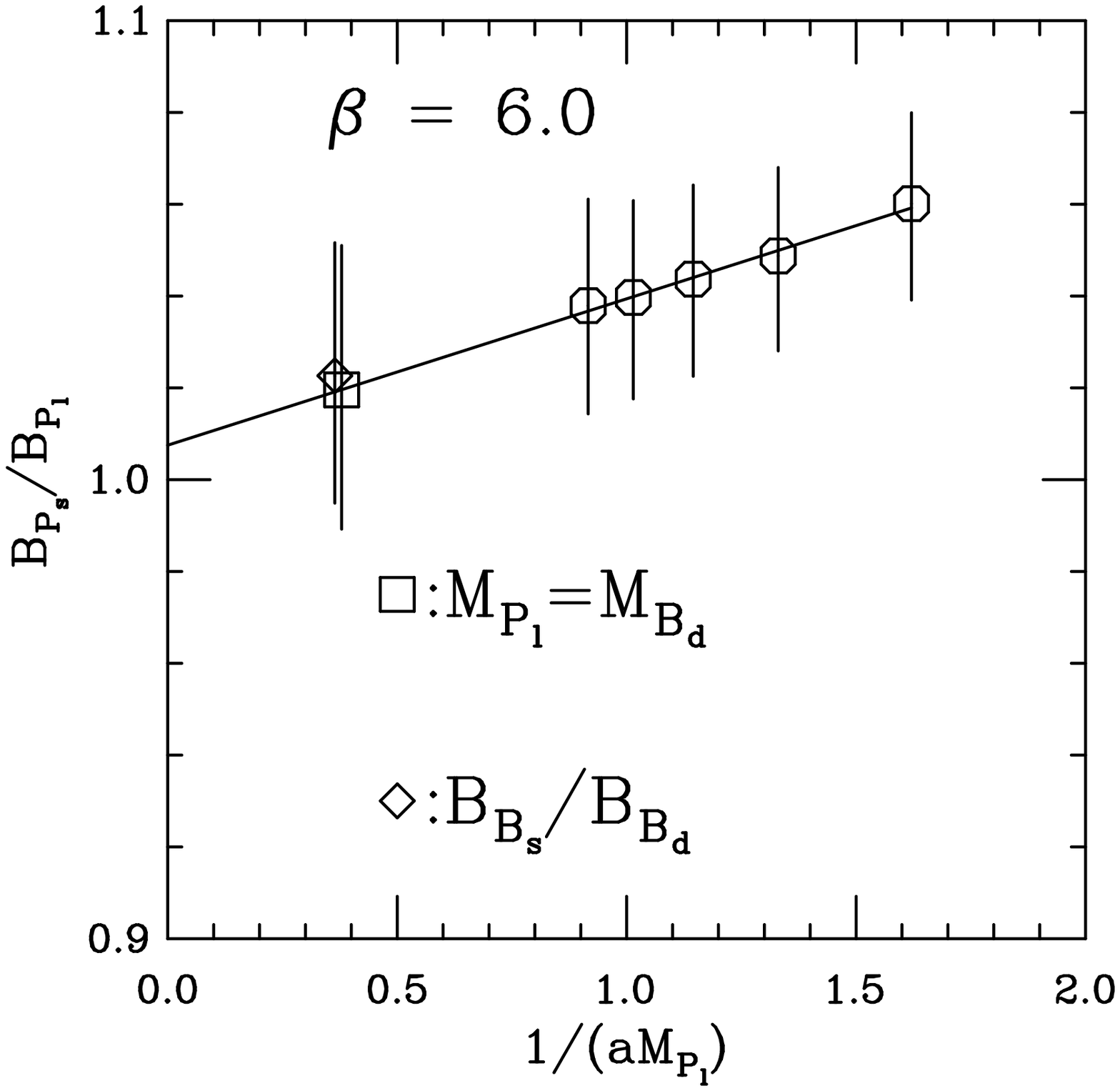}
\end{tabular}}
\caption{\label{fig:Bvshqmsu3}
Lattice results for $X(M_{P_l}) = B_{P_s}/B_{P_l}$ 
versus $1/(aM_{P_l})$ at $\beta=6.2$ and 6.0. The solid line
is a fit to the linear part of the heavy-quark-mass dependence given
in \eq{eq:hqscal}.}
\end{figure}

In \figs{fig:Bvshqm}{fig:phimatvshqm} we display our results for
$\Phi_B^q(M_{P_l})$ and $\Phi^q_{\Delta F=2}(M_{P_l})$,
constructed from $B_{P_q}(M_{P_l})$ and $\la\bar
P_q|\cO^{\Delta F=2}_q(M_{P_l})| P_q\ra$ renormalized at
$M_{P_l}$ in the NDR-$\msbar$ scheme, as functions of inverse
heavy-meson mass at $\beta=6.0$ and 6.2 and for $q=s$ and $l$. In
\fig{fig:phifvshqm}, we plot $\Phi_f^q(M_{P_l})$ versus
$1/(aM_{P_l})$.  Finally, in
\figss{fig:Bvshqmsu3}{fig:phidf2vshqmsu3}{fig:fvshqmsu3} we plot the
extrapolations of the corresponding $SU(3)$-breaking ratios.  The fit
parameters of the heavy-quark-mass scaling of the various quantities
studied are summarized in \tab{tab:hqfitparams}. While the
heavy-quark-mass dependence of most quantities is mild, that of
$\Phi_f^q(M_{P_l})$ and especially $\Phi^{q}_{\Delta F=2}(M_{P_l})$ is
quite severe.

\medskip

\begin{figure}
\centerline{\begin{tabular}{cc}
\epsfxsize=0.48\textwidth\epsffile{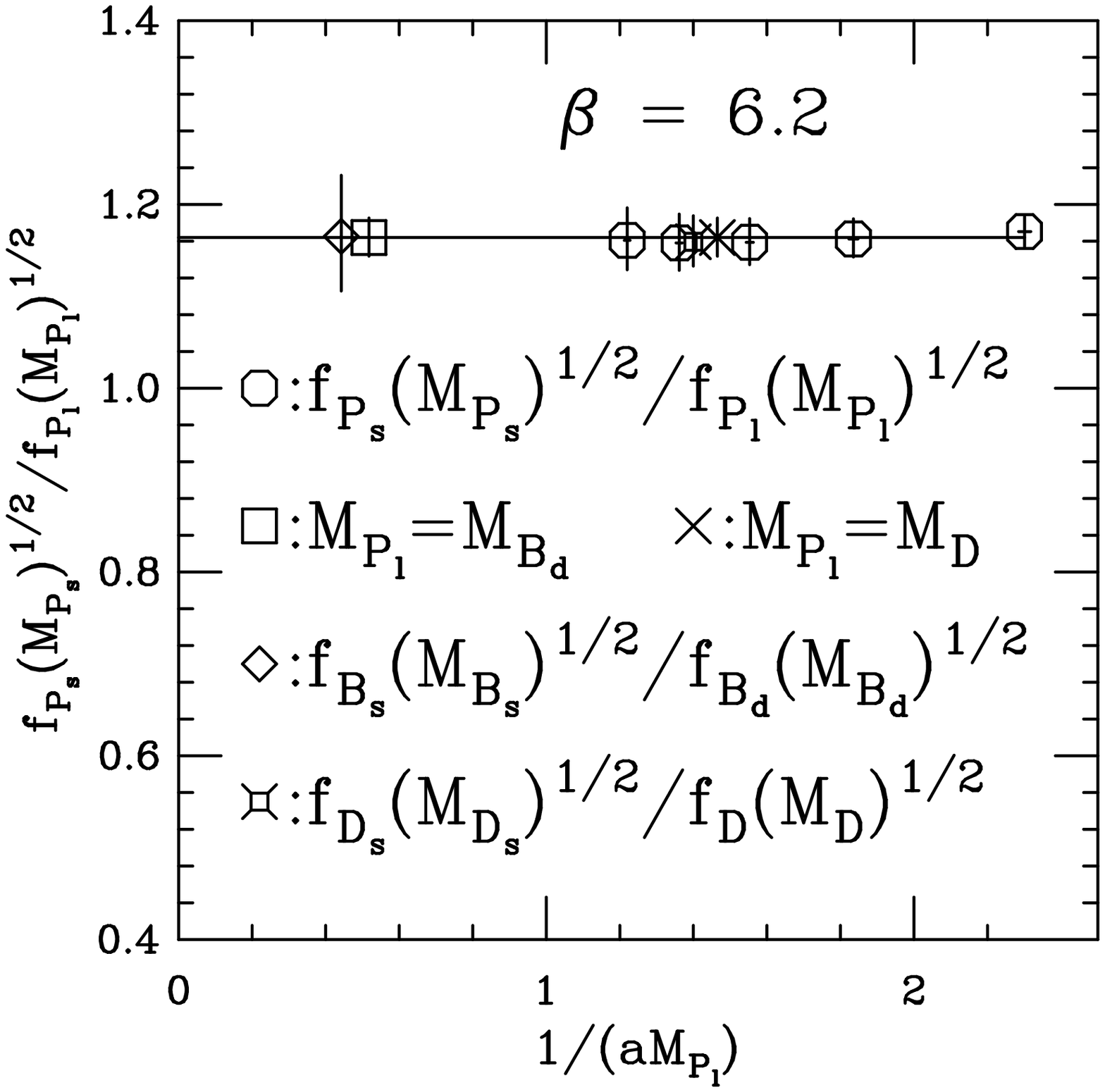}&
\epsfxsize=0.48\textwidth\epsffile{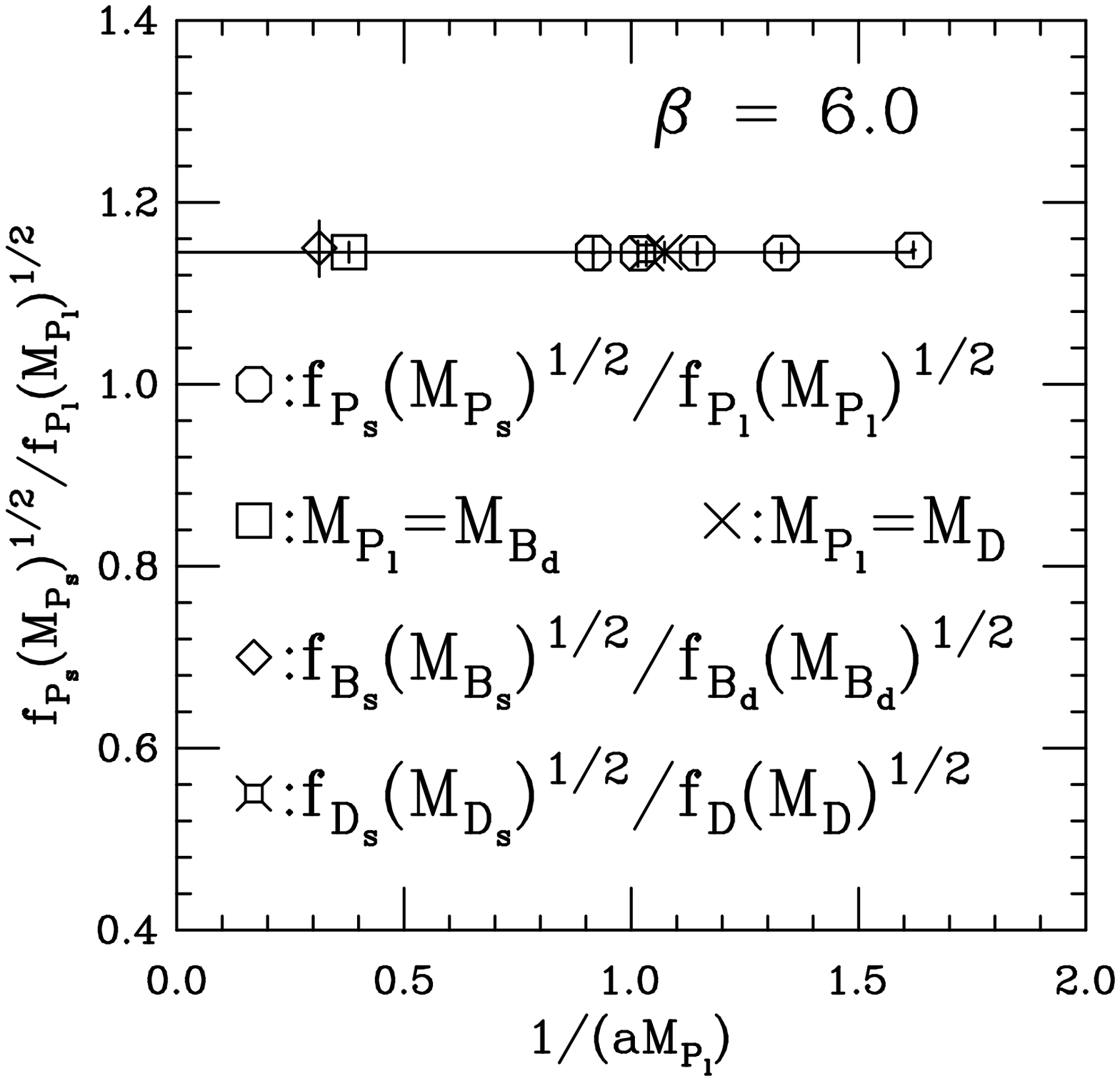}
\end{tabular}}
\caption{\label{fig:fvshqmsu3} Lattice results for
$X(M_{P_l}) = f_{P_s}/f_{P_l}\times\sqrt{M_{P_s}/M_{P_l}}$ 
versus $1/(aM_{P_l})$ at
$\beta=6.2$ and 6.0. The solid line is a fit to the constant part of
the heavy-quark-mass dependence given in \eq{eq:hqscal}.}
\end{figure}

We extrapolate $SU(3)$-breaking ratios directly in heavy-quark mass
because the mass dependence cancels partially between numerator and
denominator, making the extrapolation less pronounced and, thus, more
reliable. This is especially visible for $f_{P_s}/f_{P_l}$
(\fig{fig:fvshqmsu3}), where the extrapolation of the ratio yields
much smaller uncertainties than the ratio of the extrapolations.  In
all cases, the extrapolation of the ratio is in excellent agreement
with the ratio of extrapolations.  It is interesting to note, also,
that the heavy-quark-mass dependence of $f_{P_s}/f_{P_l}$ appears to
be the same as that of $\sqrt{M_{P_l}/M_{P_s}}$.

\section{Systematic uncertainties}
\label{sec:results}

Our main results at the two values of lattice spacing are summarized
in \tabss{tab:results}{tab:resultscont}{tab:resultscont2}.  In these
tables, the first error on each quantity is statistical.  The
remaining uncertainties are systematic and we discuss them now.

\subsection{Discretization errors}
\label{sec:discerr}

Ideally, one would extrapolate all computed quantities to the
continuum limit, where discretization errors vanish. With two values
of the lattice spacing, however, this is not possible. We must
therefore use the information that we have to estimate the uncertainty
associated with residual discretization effects.

\medskip

In \tab{tab:results}, results for the decay constants display some
dependence on lattice spacing. This suggests that discretization
errors for these quantities may be important.  The leading
discretization errors with the mean-field-improved SW action are {\it
formally} of $\ord{\alpha_sa}$, as they are for the tree-level
improved SW action. Subleading errors begin at
$\ord{\alpha_s^2a}$~\footnote{$\ord{\alpha_s^2a}$ errors, as well as
all errors proportional to $a$, are absent in non-perturbatively
$\ord{a}$-improved calculations.} and $\ord{a^2}$. To estimate these
leading and subleading errors, we consider the following variations in our
procedure.  $\ord{\alpha_sa}$ improvement of the axial current
requires one to include the effect of the $a\partial_{\mu}P$
counterterm through the replacement ($P$ is the pseudoscalar density)
\beq
 A_{\mu}\to A_{\mu} + c_{A}a\partial_{\mu}P
\ ,
\label{eq:cacontrib}\eeq
as well as to rescale the quark fields as
\beq
\psi \to (1 + \frac{b_{A}}{2}
 am_\psi)\psi
\label{eq:qnorm}\ ,\eeq
with both $c_A$ and $b_A$ evaluated at one loop
\cite{Luscher:1996sc,Luscher:1996vw}.  From a comparison of
results obtained with $c_A$ and $b_A$ set to their tree-level values
($c_A=0$ and $b_A=1$) to those obtained with $c_A$ and $b_A$ evaluated
at one loop, we can estimate the effect of $\ord{\alpha_sa}$
discretization errors. We do not use the one-loop results as central
values for the decay constants to be consistent with our determination
of the $B$ parameters. Indeed, $\ord{\alpha_sa}$ improvement of the
four-quark operators would require one to consider the mixing of these
operators with operators of dimension seven, which is beyond the scope
of the present paper.

\medskip

As already mentioned, to subtract higher-order discretization effects,
we use the EKM normalization of \eq{eq:EKLM}. Thus we define the
one-loop variation in the normalization of the quark fields through
\beq
\psi \to \frac{\sqrt{1 +  a\bar m_\psi}}{1 +  a\bar m_\psi/2}(1 + 
\frac{b_{A}^{\mathrm{1-loop}}}{2}
 a\bar m_\psi)\psi
\ ,
\label{eq:bacontrib}\eeq
where, using the results of \cite{Sint:1997jx}, we find
$b_{A}^{\mathrm{1-loop}}=1+(\alpha_{s}/4\pi)[24.03+X]=1+
10.87(\alpha_{s}/4\pi)$ for $\csw=1$. $c_A^{\mathrm{1-loop}}$ is
given by $-1.20(\alpha_{s}/4\pi)$~\cite{Sint:1997jx}. We find that
the replacements of \eqs{eq:cacontrib}{eq:bacontrib} have opposite
effects. The former lowers the decay constants while the latter
increases their values. For the one-loop values of $c_A$ and $b_A$, the
cancellation is rather good and the resulting one-loop versus
tree-level variation is certainly an underestimate of the residual
discretization errors proportional to $a$. To get a more realistic
estimate, we consider the variation brought about by each replacement
separately. These variations are shown as the second error on the
decay constants and their $SU(3)$-breaking ratios, $f_{D_s}/f_D$ and
$f_{B_s}/f_{B}$. We take the largest of the two to be a measure of
residual discretization errors proportional to $a$.~\footnote{The
largest variation is the one of \eq{eq:bacontrib}.}

\medskip

To get a handle on errors proportional to higher powers of $a$, we
consider the result of using the tree-level, quark-field normalization
of \eq{eq:qnorm} with $b_A=1$, instead of the EKM normalization of
\eq{eq:EKLM}. These two normalizations differ at $\ord{a^2}$ and we
take the resulting variation to be a measure of these additional
discretization errors. This variation is shown as the third error on
the decay constants and their $SU(3)$-breaking ratios, $f_{D_s}/f_D$
and $f_{B_s}/f_{B}$.

\medskip

The symmetric discretization error that enters the systematic error
in the final results of \eq{eq:decayconstants} is obtained by
combining in quadrature our estimates of the residual discretization
uncertainties proportional to $a$ and of the uncertainties
proportional to higher powers of $a$. While these two uncertainties
are comparable at $\beta=6.2$, the latter are significantly larger at
$\beta=6.0$ in the $b$-quark sector.

\medskip

A similar estimate of discretization errors can be carried out for the
four-quark matrix elements and their $B$ parameters. However, as we
have already mentioned a full quantification of $\ord{\alpha_sa}$
effects for these quantities is beyond the scope of this paper. In
fact, many discretization effects, such as those associated with the
normalization of quark fields, cancel or partially cancel in the
ratios of matrix elements used to define the $B$ parameters and
$SU(3)$-breaking ratios. Furthermore, in \tab{tab:resultscont},
results for $B$ parameters and $SU(3)$-breaking ratios exhibit
very little lattice-spacing dependence, supporting the idea that
discretization errors for these quantities are small. Thus, we assume
that their statistical uncertainties encompass possible residual
discretization errors.  For the quantities in \tab{tab:resultscont2},
however, which are obtained using the decay constants, we take into
account the discretization errors on these constants.

\subsection{Matching uncertainties}

As already indicated in \sec{sec:matching}, to estimate the systematic
errors arising from the perturbative matching of the various
quantities we compute, we vary the scale, $q^*$, at which
$\alpha_{\msbartiny}$ is evaluated, in the range between $1/a$ and
$\pi/a$, and compare this with the result obtained for $q^*=2/a$.  We also
consider the variation coming from computing $Z_{11}(M_{P_l})$,
$Z_{1i}$ and $Z_A$ with the constant $\csw$ set to its
mean-field-improved value instead of 1, keeping $q^*=2/a$
fixed. $SU(3)$-breaking ratios of decay constants are not affected by
these variations while those of the $B$ parameters are not significantly
so. These variations are reflected in the fourth error in
\tabss{tab:results}{tab:resultscont}{tab:resultscont2}.

\subsection{Heavy-quark-mass extrapolations}

\begin{figure}
\centerline{\begin{tabular}{cc}
\epsfxsize=0.48\textwidth\epsffile{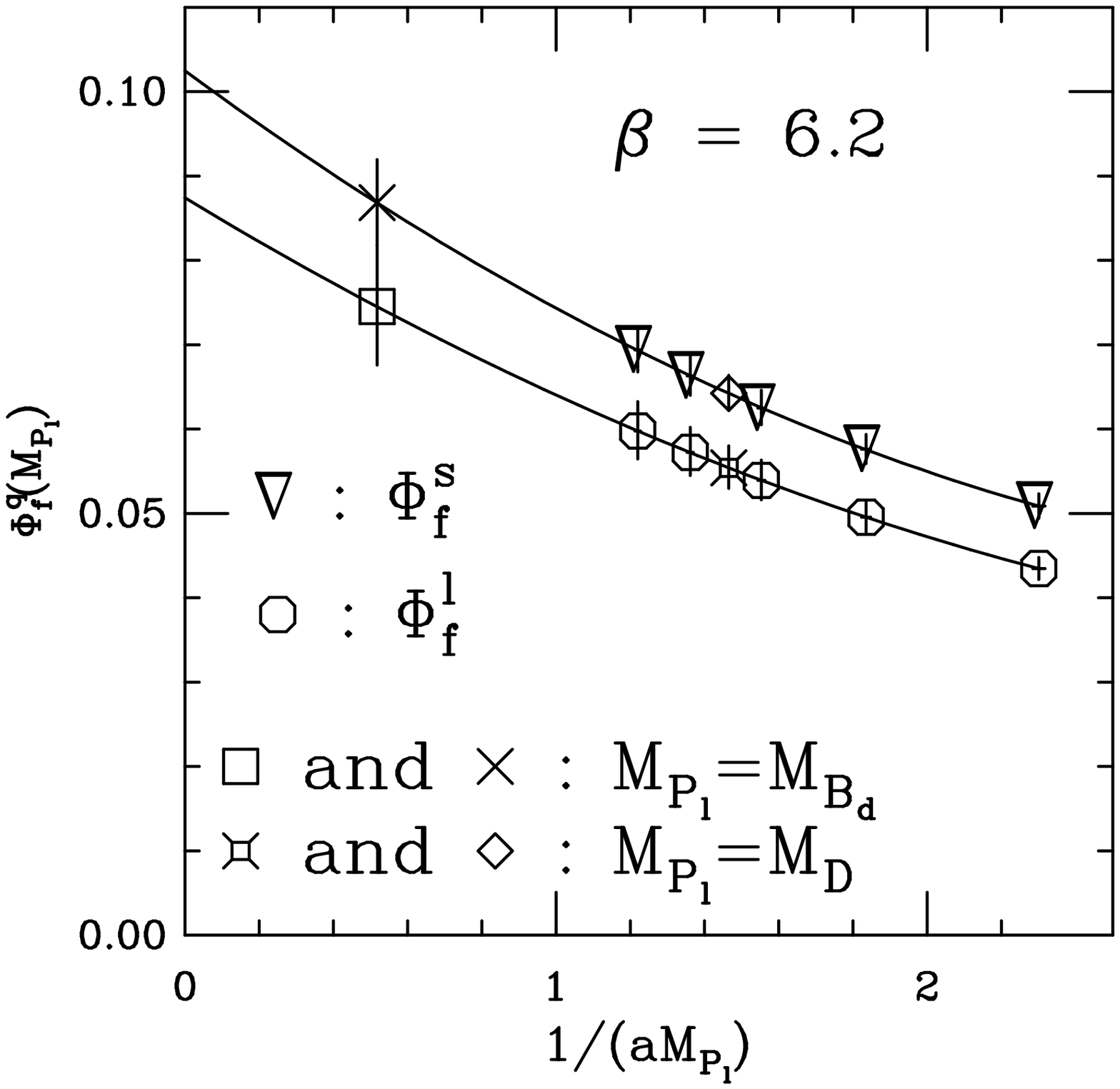}&
\epsfxsize=0.48\textwidth\epsffile{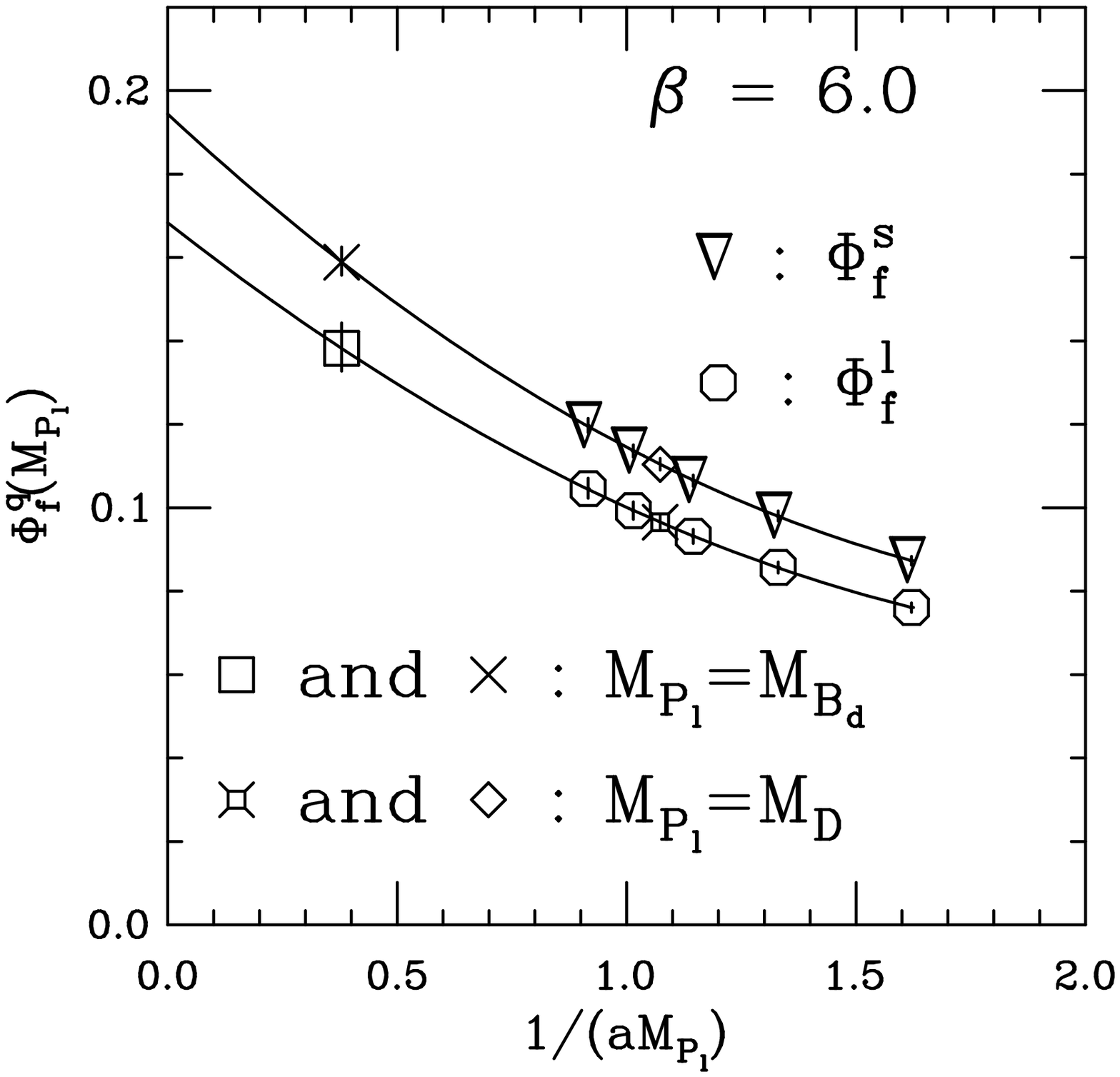}
\end{tabular}}
\caption{\label{fig:phifvshqm}
Lattice results for 
$X(M_{P_l}) = \Phi_f^q(M_{P_l})$ versus $1/(aM_{P_l})$ 
at $\beta=6.2$ and 6.0
and for $q=s$ and $l$. The solid curves
are quadratic fits of the heavy-quark-mass dependence given
in \eq{eq:hqscal}.}
\end{figure}

As shown in \fig{fig:phifvshqm}, the decay constants have a pronounced
extrapolation in heavy-quark mass, and the term quadratic in
$1/(aM_{P_l})$ on the right-habd side 
of \eq{eq:hqscal} contributes significantly.
To quantify the systematic error associated with this
extra\-polation--the fifth error on the decay con\-stants--we perform
a fit of the heaviest three points in \fig{fig:phifvshqm} to the 
right-hand side
of \eq{eq:hqscal}, without the quadratic term. For the
$SU(3)$-breaking ratios of decay constants, we perform a constant fit
to these same three points. These uncertainties are propagated to the
results of \tab{tab:resultscont2}.

\medskip

\fig{fig:Bvshqm} indicates that the heavy-quark-mass dependence of the
$B$ parameters and their $SU(3)$ breaking ratio is mild and to very
good approximation linear.  We have verified that a linear fit to the
three heaviest points gives results that are well within the errors
bars of the fit to all five points at $\beta=6.0$. We assume that the
same would be true at $\beta=6.2$ if we also had five heavy quarks, as
there is no evidence for curvature on the three points that we have.

\medskip

\begin{figure}
\centerline{\begin{tabular}{cc}
\epsfxsize=0.48\textwidth\epsffile{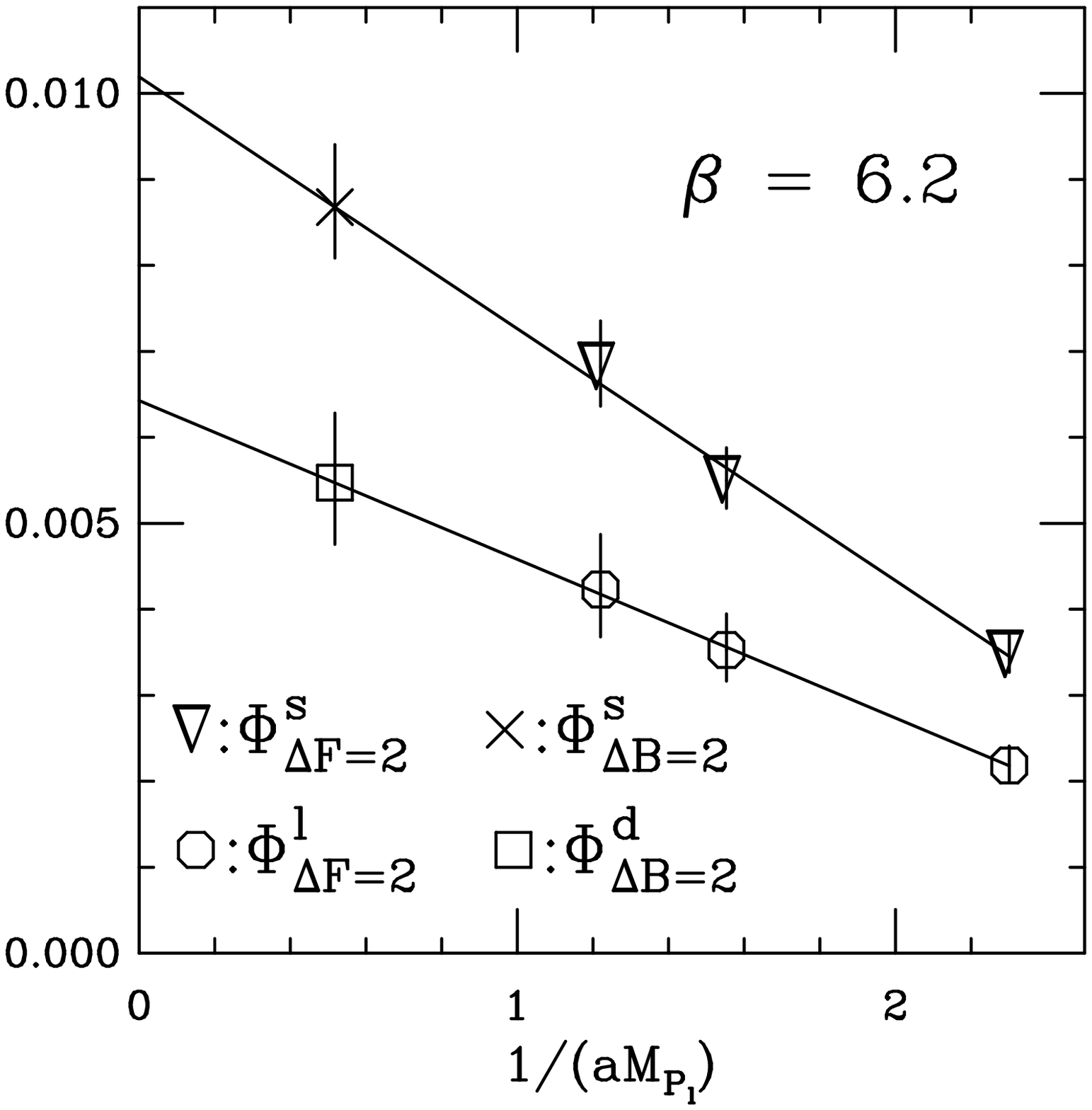}&
\epsfxsize=0.48\textwidth\epsffile{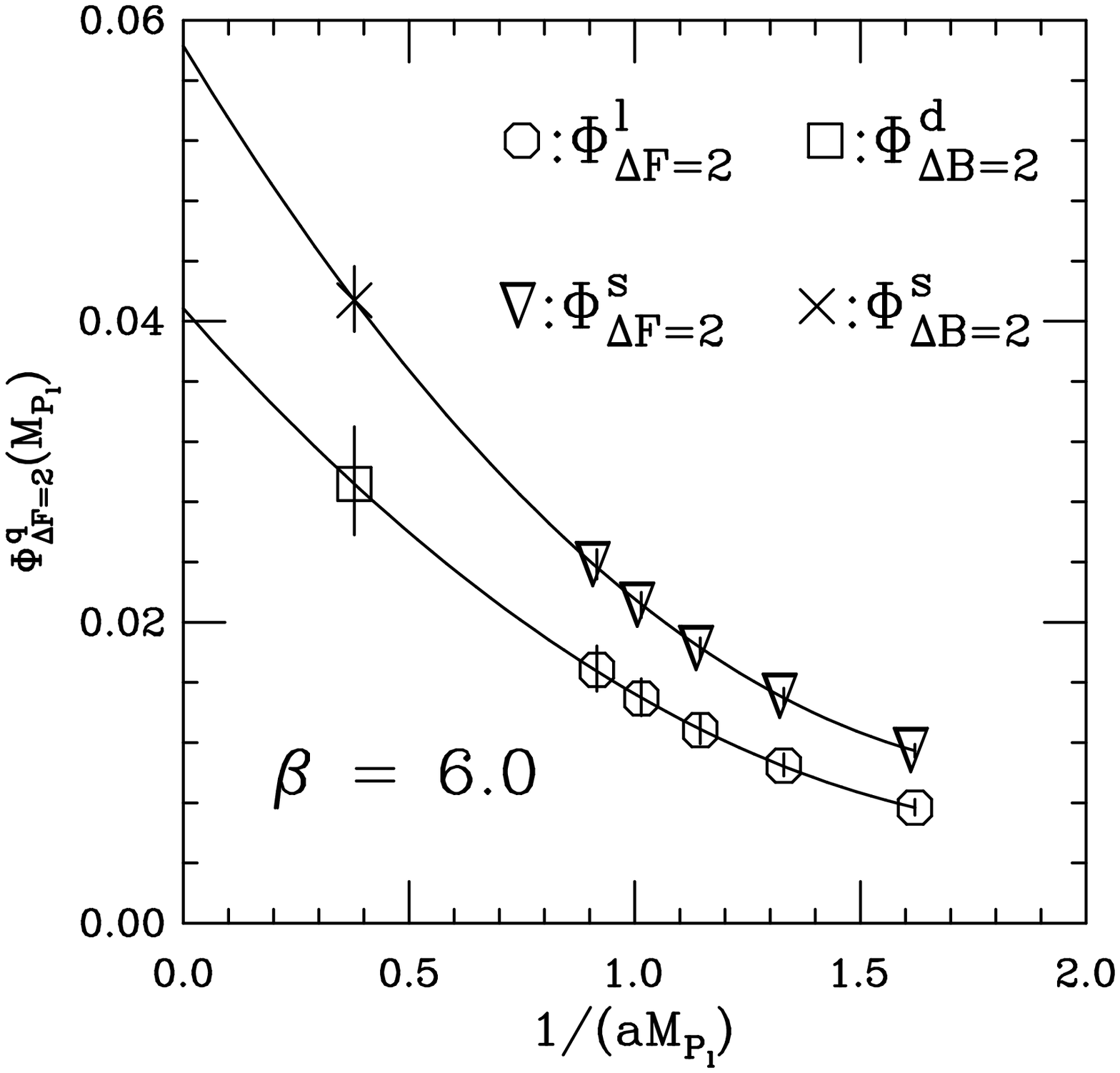}
\end{tabular}}
\caption{\label{fig:phimatvshqm}
Lattice results for $X(M_{P_l}) = \Phi_{\Delta F=2}^q(M_{P_l})$ 
versus $1/(aM_{P_l})$ at $\beta=6.2$ and 6.0
and for $q=s$ and $l$. The solid curves
are linear and quadratic fits of the heavy-quark-mass dependence given
in \eq{eq:hqscal} at 
$\beta=6.2$ and $\beta=6.0$, respectively.}
\end{figure}

The $\Delta F = 2$ matrix elements have a very pronounced dependence
on heavy-quark mass, as seen in \fig{fig:phimatvshqm}. Since we are not
reporting results for the four-quark matrix elements themselves, we do
not quantify the systematic errors associated with their
determination.  

\medskip

\begin{figure}
\centerline{\begin{tabular}{cc}
\epsfxsize=0.48\textwidth\epsffile{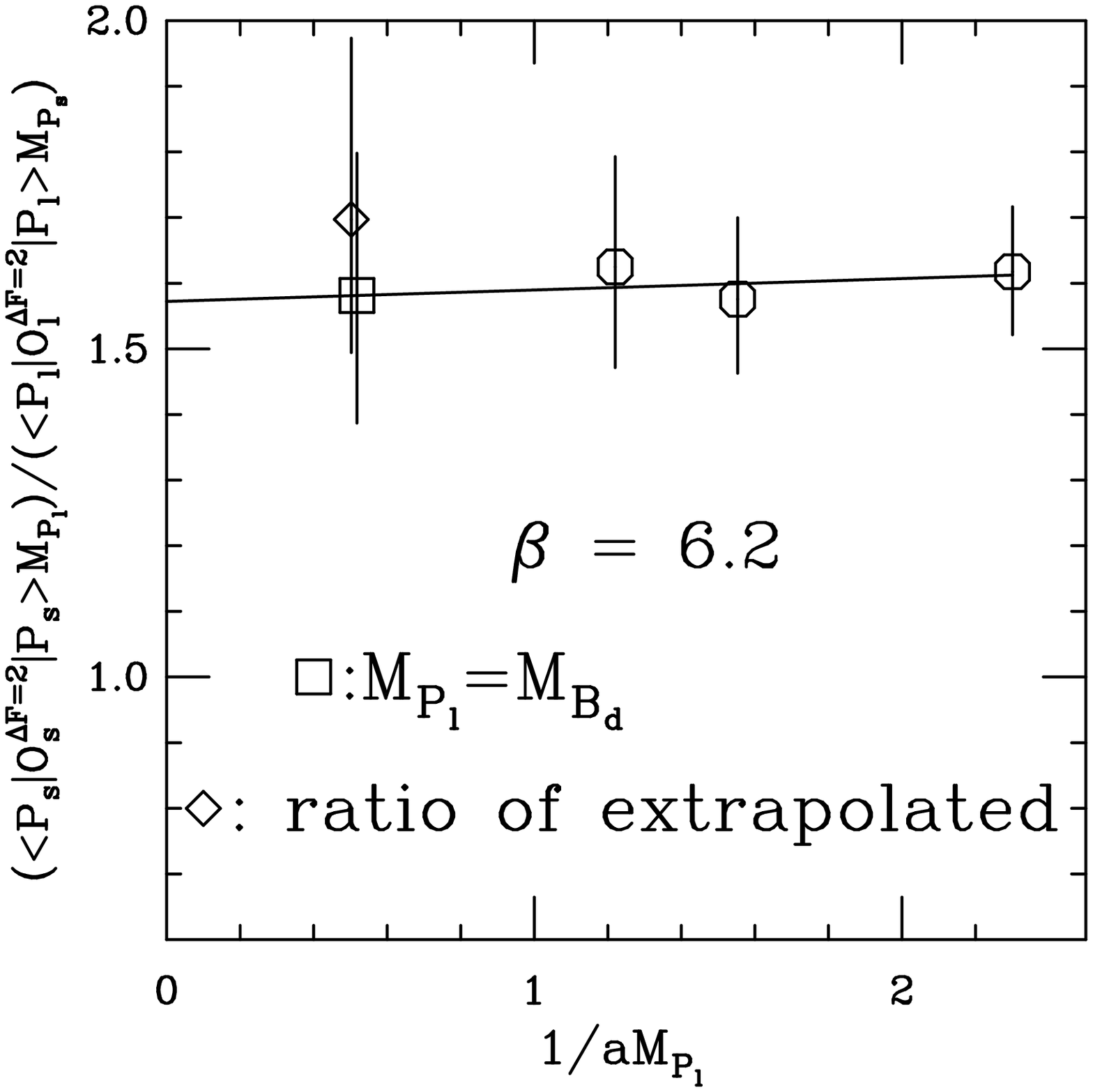}&
\epsfxsize=0.48\textwidth\epsffile{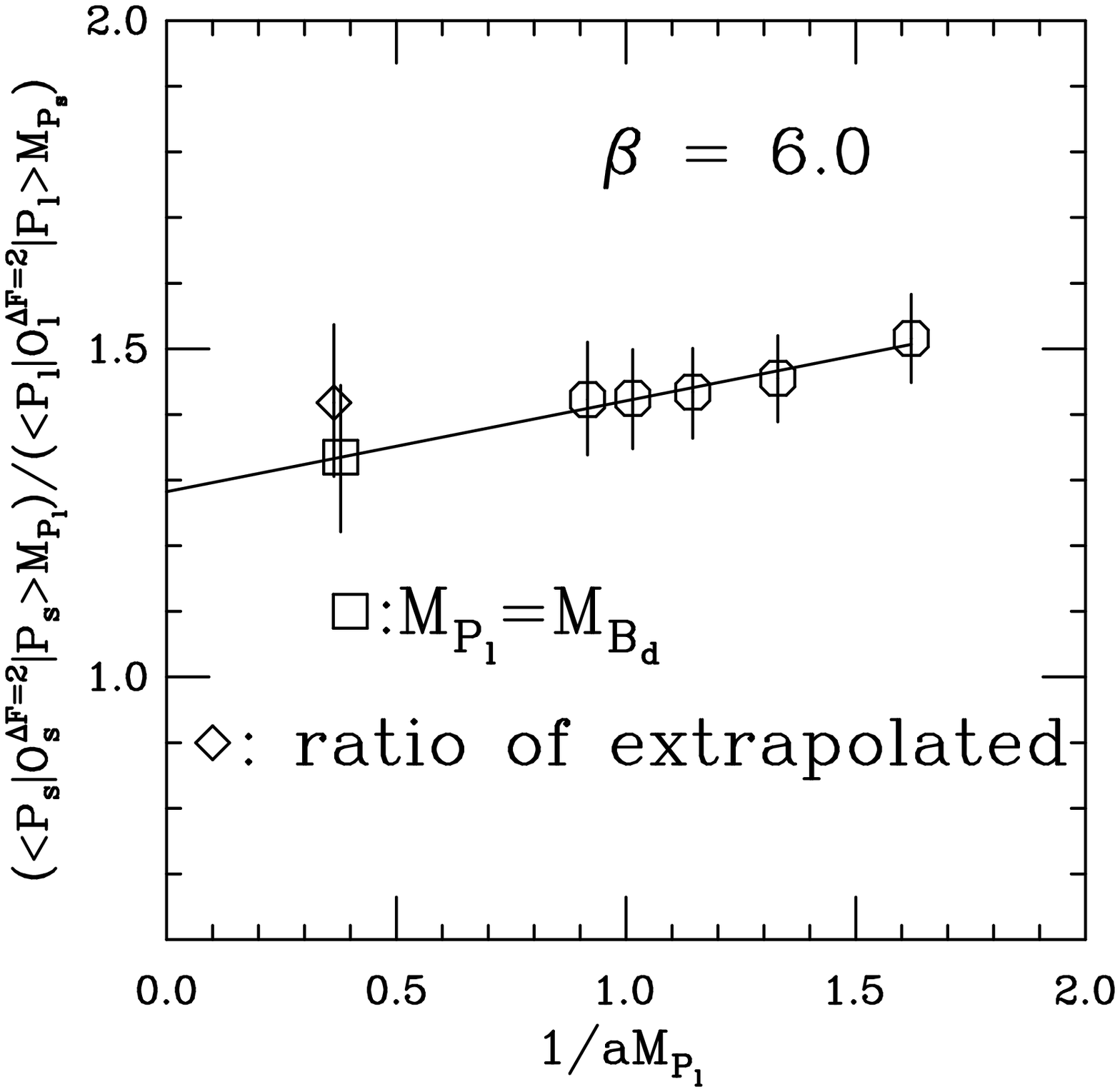}
\end{tabular}}
\caption{\label{fig:phidf2vshqmsu3} Lattice results for $X(M_{P_l}) 
=(\la\bar 
P_s|\cO_s^{\Delta F=2}| P_s\ra/ \la\bar
P_l|\cO_l^{\Delta F=2}| P_l\ra)\times(M_{P_l}/M_{P_s})$ versus
$1/(aM_{P_l})$ at $\beta=6.2$ and 6.0. The solid line is a fit to the
linear part of the heavy-quark-mass dependence given in \eq{eq:hqscal}.}
\end{figure}

One may worry that we have only three heavy quarks at $\beta=6.2$ in
our calculation of $\Delta F=2$ matrix elements. However, as we have
just seen, the heavy-quark scaling of the $B$ parameters and their
$SU(3)$-breaking ratio is mild and displays no evidence for
curvature. This is confirmed by the behavior at $\beta=6.0$ where we
have five heavy quarks. Thus, we believe that our results for
$B$ parameters and the derived quantities of \tab{tab:resultscont2} at
$\beta=6.2$ are reliable. The situation is certainly not as favorable
for the $\Delta F=2$ matrix elements themselves. For those, there is
evidence for curvature and our three points at $\beta=6.2$ can be
thought to yield only a rough estimate. Thus, we do not attempt to
give a final result for these matrix elements from this procedure. In
fact, this strong mass de\-pen\-dence is one of the problems that
makes a reliable determination of $r_{sd}$, from the ratio of
individually calculated $\la\bar B_q|\cO^{\Delta
B=2}_q|B_q\ra$, $q=s,d$, difficult. The extrapolation of $r_{sd}$
itself, on the other hand, is much milder and the curvature is much
reduced. Thus, we extrapolate it linearly and verify, at $\beta=6.0$
where we have enough points, that the result of a quadratic fit,
$r_{sd}^{\mathrm{direct}}=1.50(17)$, is entirely
compatible.~\footnote{To estimate the heavy-quark extrapolation
error at $\beta=6.0$, we consider the variation due to the removal
of the lightest two points from the linear extrapolation of Fig.
\ref{fig:phidf2vshqmsu3}.} In any event, the final value of $r_{sd}$ that
we quote is that given by the ``indirect'' method, where none of this
is a problem.

\medskip

Another concern may be that our lightest heavy quark is too light to
be in the heavy quark scaling regime.  However, in the extrapolations used 
to obtain the results of Eqs. (\ref{eq:mainresults}) 
and (\ref{eq:decayconstants}), the points corresponding to
this quark are consistent with the smooth curves determined by the 
other points.
Furthermore, our heaviest
quarks are as massive as those in other relativistic calculations (see
for instance Ref. \cite{Becirevic:2000nv}). Thus, we are not distorting the
heavy-quark extrapolations by including these lighter points, nor are
we missing information on the heavier-quark end. Finally, as described
above, we include in our errors, the variation obtained by ignoring
our lightest two points, where appropriate.

\medskip

Ideally, one would have continuum extrapolations of results such as
ours and of the same quantities computed in the static limit
(corresponding to an infinite-mass heavy quark). Results for the $b$
would then be obtained by an interpolation in heavy-quark mass instead
of by extrapolation, as they are here. We leave such studies for the
future.

\subsection{Uncertainties in the determination of the lattice spacing}

In quenched calculations, the value of the lattice spacing varies
significantly with the quantity used to set the scale.  This variation
may be due, in part, to quenching effects, as well as any other
systematic uncertainty that affects the quantity used to set the
scale. In this paper, we determine the lattice spacing from $f_K$ and
$m_K$, as described in \sec{sec:chiral}.  We then vary the inverse
lattice spacing, $a^{-1}$, by $\pm 7\%$.  This range covers the
variations observed in the determination of the scale from gluonic or
light-hadron spectral quantities, and with the same action and parameters
as we use \cite{Bowler:1999ae}~\footnote{Some of the baryons considered in
Ref. \cite{Bowler:1999ae} would give lattice spacings outside our $\pm
7\%$. However, these particles are more susceptible, than the
particles we are studying here, to systematic effects such as those
associated with the finite volume of the lattice.}, as well as the
variation due to the uncertainty in the perturbative determination of
$Z_A$.

\medskip

Uncertainties in the lattice spacing will obviously affect the
determination of the decay constants and the dimensionful quantities
derived from them.  They will also slightly change the length of the
heavy-quark-mass extrapolations
(\figss{fig:phifvshqm}{fig:Bvshqm}{fig:phimatvshqm}).  Furthermore,
they induce a variation of order $\pm 15\%$ in $a\tilde m_s$,
which we obtain from the mass of the kaon, and therefore affect all
quantities that depend on this mass.

\medskip

In practice, we find that the variation of the lattice spacing
discussed above does not induce a significant change in the
$B$ parameters.  However, it does affect all the decay constants and the
corresponding $SU(3)$-breaking ratios as well as the quantities in
\tab{tab:resultscont2} which are obtained from these constants.  All of these
observations are reflected in the sixth error on the quantities in
\tabss{tab:results}{tab:resultscont}{tab:resultscont2}.

\subsection{Determination of the light-quark masses}

Another source of systematic error is the uncertainty in the
determination of the light-quark masses. In the previous section we
saw that changing the lattice spacing, while keeping fixed the
physical quantity used to set the strange-quark mass, induced
approximatively a $\pm 15\%$ variation in $a\tilde m_s$.  One can also
imagine doing the reverse, i.e. holding the lattice spacing fixed
while varying the observable used to set the strange-quark mass.  For
instance, we could have used the $K^*$ or $\phi$ meson masses to fix
$a\tilde m_s$, instead of $m_K$. Let us denote the resulting values of
the strange quark mass by $a\tilde m_s(m_{K^*})$, etc. Due to
quenching and other systematic effects, the values obtained may
differ.

To estimate these differences, we turn to the literature. In 
Ref. \cite{Bhattacharya:1998ht}, where the determination of quark
masses with different fermionic actions is reviewed thoroughly,
Bhattacharya and Gupta conclude that for mean-field-improved SW
actions such as ours, $a\tilde m_s(m_\phi)/a\tilde m_s(m_K)\simeq
1.2$, a statement that they find depends very little on the gauge
coupling $\beta$, at least in a range that covers our two simulations.
This result is obtained with the lattice spacing fixed by the mass of
the $\rho$ meson. With the lattice spacing set in this way, it is
straightforward to show that $a\tilde m_s(m_{K^*})\simeq a\tilde
m_s(m_\phi)$. This follows from the observed linear behavior of the
light vector meson masses and the fact that $m_\rho\simeq
2m_{K^*}-m_\phi$ in nature.  The picture changes when the value of the
lattice spacing differs from the one given by the mass of the $\rho$,
$a(m_\rho)$. In the present case, however, the values of the lattice
spacing that we use appear to be consistent, within errors, with the
values of $a(m_\rho)$ that can be inferred from Ref. 
\cite{Bhattacharya:1998ht}. Thus, we consider that 
the 20\% upward variation in
$a\tilde m_s$ described in Ref. \cite{Bhattacharya:1998ht} is a reasonable
estimate of the uncertainty associated with the different possible
choices of an observable to fix this mass. The changes that this variation
induces in our results are
reflected in the seventh error on
the quantities in \tabss{tab:results}{tab:resultscont}{tab:resultscont2}.

\subsection{Quenching errors}
\label{sec:quench}

Quenching effects for the quantities of interest here have been
studied using quenched chiral perturbation theory
\cite{Sharpe:1996qp}. They are typically a few percent for the
$B$ parameters if the theory's couplings are constrained by
large-$N_c$ arguments and by the reasonable range of Ref.
\cite{Sharpe:1996qp}, and larger outside these ranges.  Recent results
for the decay constants, obtained with two flavors of dynamical
quarks ($n_f=2$), show little variation in $f_{B_s}/f_{B}$ compared to
its quenched value
\cite{Bernard:1999nv,Collins:1999ff,Khan:2000eg,Bernard:2000unq}. The
authors of \cite{Khan:2000eg} find that this ratio is enhanced by
$(5\pm 3)\%$ in their calculation with light dynamical quarks. Thus,
if quenching effects on the $B$ parameters are small, commensurate
variations on $r_{sd}$ and $\xi$ are expected. For the decay constants
of $B$ ($B_{s}$) mesons, quenching effects appear to be significant
\cite{Bernard:1999nv,Collins:1999ff,Khan:2000eg,Bernard:2000unq}. For
instance, the authors of Ref. \cite{Khan:2000eg} find that these decay
constants are enhanced by 11\% (14\%) when light-quark loops are
included, with a statistical significance of 2 to 3 standard
deviations. For the $D$ ($D_{s}$) mesons, the effect is of 3\% (7\%) and
consistent with zero within roughly one standard deviation. These
latter results suggest that quenching errors, at least on the
$B$-meson decays constants, may be reduced by normalizing these
constants with $f_{D_s}$. The reduction of quenching effects is 
about 7\% and the remaining effects become consistent with zero. To
the extent that quenching errors on the $B$ parameters are negligible,
all of these considerations carry over to the quantities in
\tab{tab:resultscont2} which are proportional to decay constants. It
should be remembered, however, that the real world has a third
dynamical light quark and the effects discussed above may be
amplified.

\medskip

A thorough estimate of quenching effects for all of the quantities
that we calculate would require a dedicated unquenched simulation,
which is beyond the scope of this paper. Therefore, we do not attempt
to quantify these effects specifically. Nevertheless, as we mentioned
in the two previous sections, the uncertainties in the lattice scale
and in the strange quark mass are, at least in part, quenching
effects.

\section{Final results and discussion}
\label{final}

Because of the excellent consistency of the results for $B_{B_s}$,
$B_{B_d}$, $B_{B_s}/B_{B_d}$ and $f_{B_s}/f_{B_d}$
at the two lattice spacings and because we cannot perform a continuum
extrapolation with only two points, we take the results from the finer
lattice at $\beta=6.2$, which should have smaller discretization
and matching uncertainties, to be our best estimates.

\medskip

For the $SU(3)$-breaking ratio $r_{sd}$, we also
have a choice between the ``direct'' and ``indirect'' approaches
described in the Introduction. While both methods give results that
are compatible at the two values of the lattice spacing, the ``direct''
method leads to larger variations with lattice spacing and
significantly larger statistical errors.  Furthermore, as discussed in
\secs{sec:chiral}{sec:hq}, the light- and heavy-quark-mass
extrapolations are better behaved in the ``indirect'' method.
Therefore, we take $r_{sd}$ obtained with the ``indirect'' method at
$\beta=6.2$ as our best estimate for this quantity.

\medskip

A summary of our results for quantities directly relevant for $B{-}\bar
B$ mixing is given in \eq{eq:mainresults}. These results are
compatible with previous calculations of some or all of these
quantities, which were performed using less improved relativistic
fermion actions \cite{Bernard:1998dg,Aoki:1995jq}, as well as with the
recent calculation of Ref. \cite{Becirevic:2000nv}, which makes use of a
non-perturbatively, $\ord{a}$-improved, Sheikholeslami-Wohlert
action. While the decay constants in Ref. \cite{Becirevic:2000nv} are
non-perturbatively improved,~\footnote{Almost: the authors actually
use the perturbative value of $b_A$.} $B$ parameters and four-quark
matrix elements are not. Thus, for those quantities, the
discretization accuracy of that calculation is formally the same as
ours. Moreover, the authors of Ref. \cite{Becirevic:2000nv} do not
investigate cutoff dependence, as we do here with our two lattice
spacings. Our results are also consistent, once systematic errors are
taken into account, with the NRQCD results of Refs.
\cite{Hashimoto:2000eh,Yamada:2000ym}, whose $B$ parameters are 
$7\sim 10\%$ smaller than ours.

\medskip

For the decay constants and quantities proportional to them, the
situation is less favorable than for $B$ parameters and 
$SU(3)$-breaking ratios.  We do observe a
two-statistical-standard-deviation dependence on lattice spacing,
indicating that discretization errors are more important here. 
We quantify these discretization effects, as described in
\sec{sec:discerr}. The corresponding uncertainty at $\beta=6.0$ is
large enough to bridge the gap between the results for $B$-meson decay
constants at the two lattice spacings. For $D$-meson decay constants,
agreement requires that one also take into account the statistical
error on the $\beta=6.2$ results. So we take as our best estimates the
results from the finer lattice, which in principle have smaller
discretization and matching uncertainties, confident that our errors
are a reasonable estimate of the uncertainty associated with this
fixed lattice spacing calculation. In the future, though, when $f_{D_s}$
is accurately measured experimentally, it will be advantageous to
consider the values of these quantities in units of this decay
constant. With this normalization, the discrepencies between the results at
$\beta=6.2$ and $6.0$, as well as the size of systematic (and
statistical) errors, are significantly reduced, as can be seen in
Tables \ref{tab:results} and \ref{tab:resultscont2}.  

Our results for
the decay constants are summarized in Eq. (\ref{eq:decayconstants}).
They are compatible with other recent calculations in the quenched
approximation
\cite{Allton:1997tv,El-Khadra:1998hq,AliKhan:1998df,Bernard:1998xi,
Ishikawa:1999xu,Khan:2000eg,Bernard:2000unq}, as reviewed in Refs.
\cite{Draper:1998ms,Hashimoto:1999bk,Lubicz:2000dj}.  A very recent,
non-perturbatively $\ord{a}$-improved calculation by the UKQCD
Collaboration \cite{Bowler:2000xw} yields $f_{B_s}$ and $f_B$ which
are over two statistical standard deviations higher than our results.
The authors of Ref. \cite{Bowler:2000xw} use the scale $r_{0}$
\cite{Sommer:1994ce} to set the inverse lattice spacing, $a^{-1}$,
which gives an $a^{-1}$ at the top of our range. They further use the
preliminary non-perturbative values of $b_A$ obtained in
\cite{Bhattacharya:1999uq,Bhattacharya:1999cb} and the
non-perturbative values of $c_A$ obtained in
\cite{Luscher:1997iy}.~\footnote{The authors of Ref.
\cite{Bhattacharya:1999uq,Bhattacharya:1999cb} have since finalized
their determinations of $b_A$, as well as those of $c_A$
\cite{Bhattacharya:2000pn}. The authors of Ref. \cite{Bowler:2000xw} are
currently reviewing their analysis of decay constants to incorporate,
amongst other considerations, this new information \cite{chrism}.}
Agreement with our results is recovered, nonetheless, when systematic
errors are considered. For $D$-meson decay constants, their results
agree with ours within statistical errors.

\medskip

All of our results are obtained in the quenched approximation. Some of
the error resulting from this approximation is accounted for by the
variations that the uncertainties in the lattice spacing and the
strange quark mass induce, since these uncertainties are, at least in
part, quenching effects. However, a thorough estimate of quenching
effects for all the quantities that we consider here would require a
dedicated unquenched simulation, which is beyond the scope of this
paper.

\section{Conclusion}

We have reported on high statistics, quenched lattice calculations of
matrix elements relevant for neutral meson mixing, and leptonic decays
of $B$ and $D$ mesons. We use mean-field-improved
Sheikholeslami-Wohlert actions to describe quarks and work at two
values of the lattice spacing. We have performed an extensive study of
systematic errors and we believe that our final results, presented in
\eqss{eq:mainresults}{eq:decayconstants}{eq:ddbar}, carry errors that
reflect conservatively the uncertainty associated with our fixed
lattice spacing calculations.

\medskip

Our results for neutral $B$-meson mixing are compatible with the
results of other calculations of some or all of the quantities we
consider \cite{Hashimoto:2000eh,Yamada:2000ym,Becirevic:2000nv,
Bernard:1998dg,Aoki:1995jq}, as well as with the world averages of
\cite{Wittig:1997bf,Flynn:1997ca,Draper:1998ms,Sharpe98,Lellouch:1999ir,
Hashimoto:1999bk,Lubicz:2000dj}. The same is true of our results for
the decay constants, which are compatible with other modern, quenched
determinations
\cite{Allton:1997tv,El-Khadra:1998hq,AliKhan:1998df,Bernard:1998xi,
Ishikawa:1999xu,Khan:2000eg,Bernard:2000unq} and the world averages
of \cite{Wittig:1997bf,Flynn:1997ca,Draper:1998ms,Sharpe98,
Lellouch:1999ir,Hashimoto:1999bk,Lubicz:2000dj}.

\medskip

Finally, it should be emphasized that all of these results are
obtained in the quenched approximation. They include a quenching error
only to the extent that the variations in the lattice spacing and
strange quark mass that we account for are quenching effects.  It is
worth noting that when dimensionless quantities are considered, such
as $B$ parameters, $SU(3)$-breaking ratios or quantities normalized by
$f_{D_{s}}$, a number of systemaitc uncertainties, including possibly
those associated with quenching, partially cancel.  Nevertheless, a
thorough quantification of quenching effects for neutral $B$-meson
mixing matrix elements would require a dedicated unquenched
simulation, which is beyond the scope of this paper. It is important,
however, that such a study be undertaken. The pioneering $n_f=2$
studies of decay constants
\cite{Bernard:1999nv,Collins:1999ff,Khan:2000eg,Bernard:2000unq} are a
first step in this direction.

\medskip

\begin{Large}
\begin{flushleft}
{\bf Acknowledgments}
\end{flushleft}
\end{Large}

\medskip

This work is supported by EPSRC and PPARC under Grants Nos. GR/K41663 and
GR/L299\-27.  We acknowledge the help from the UKQCD Collaboration,
especially Pablo Martinez and Peter Boyle, on generating the lattice
data.  We thank Damir Becirevic, Ken Bowler, Jonathan Flynn, Leonardo
Giusti, Richard Kenway, Vittorio Lubicz, Craig McNeile, Guido
Martinelli, Chris Maynard, Brian Pendleton, David Richards, Chris
Sachrajda and Hartmut Wittig for useful discussions.  L. L. thanks the
University of Edinburgh and C.-J. D. L.  thanks Centre de Physique
Th\'eorique of CNRS Luminy and the Theory Division of CERN for their
hospitality.  C.-J. D. L. acknowledges the financial support of the
Department of Physics and Astronomy of the University of Kentucky
under DOE EPSCor grant DE-FG05-84ER40154.

\addcontentsline{toc}{chapter}{Bibliography}
\bibliographystyle{prsty}
\bibliography{refs}

%
%
\begin{table}
\begin{center}
\caption{\label{tab:simparam}
Simulation parameters. $\csw$ is the clover coefficient.}
\begin{tabular}{ccc}
\\
\hline\hline
$\beta$ & 6.2 & 6.0\\ 
\hline
$\csw$ & 1.442 & 1.479\\
size & $24^3\times 48$ & $16^3\times 48$\\
$\#$ of configurations & 188 & 498\\
\hline\hline
\end{tabular}
\end{center}
\end{table}
\begin{table}
\begin{center}
\caption{\label{tab:qmasses} Hopping parameters, $\kappa_\psi$, of
the light ($\psi=q$) and heavy ($\psi=Q$) quarks used in the
simulations.  In brackets, we give the masses, $m_p$, of light-light
pseudoscalar mesons composed of a degenerate quark and antiquark with
hopping parameter $\kappa_q$, obtained as described in
\sec{sec:corfunc}. We also provide the masses, $M_P$, of heavy-light
pseudoscalar mesons composed of a heavy quark with hopping parameter
$\kappa_Q$ and a massless antiquark, obtained as described in
\sec{sec:corfunc} and \sec{sec:hq}. Error bars
are statistical only. The scales used to translate these masses into
MeV are $a^{-1}=2.73\gev$ at $\beta=6.2$ and $a^{-1}=2.00\gev$ at
$\beta=6.0$, as obtained in \sec{sec:chiral}.}
\begin{tabular}{cccccccc}
\\
\hline\hline
$\beta$ & $\kappa_q\,[m_p\mbox{ in MeV}]$ & $\kappa_Q\,[M_P\mbox{ in MeV}]$ \\
\hline
6.2 & 0.13640 [831(4)], 0.13710 [608(4)],
& 0.120 [2238(14)], 0.123 [2006(13)],\\
& 0.13745 [466(6)] &  0.126[1757(11)], 0.129 [1488(10)] \\
& & 0.132[1186(9)]\\
\hline
6.0 & 0.13700 [827(1)], 0.13810 [587(3)],
& 0.114 [2183(6)], 0.118 [1971(5)],\\
& 0.13856 [454(3)] & 0.122 [1746(4)],
0.126 [1503(4)],\\
& & 0.130 [1234(3)] \\
\hline\hline
\end{tabular}
\end{center}
\end{table}
\begin{table}
\begin{center}
\caption{\label{tab:alphav}
The coupling $\alpha_{\msbartiny}(q)$ at different values of 
$q$ as obtained
in the simulations at $\beta=6.0$ and 6.2.}
\begin{tabular}{cccc}
\hline\hline
$\beta$ & $\alpha_{\msbartiny}(1/a)$ &  $\alpha_{\msbartiny}(2/a)$ &  
$\alpha_{\msbartiny}(\pi/a)$ \\
\hline
6.2 & 0.1730 & 0.1402 & 0.1250 \\
6.0 & 0.1921 & 0.1522 & 0.1343 \\
\hline\hline
\end{tabular}
\end{center}
\end{table}
\begin{table}
\caption{\label{tab:calcandfittimes}
Fit ranges for the correlation functions and ratios used in this work.}
\begin{center}
\begin{tabular}{ccc}
\hline\hline
$\beta$ & 6.2 & 6.0 \\
\hline
$C_{\cO}^{FLF}/C_{AP}^{LF}C_{AP}^{LF}$&
$34 \le t_{x} \le 38$ & $33 \le t_{x} \le 37$ \\
 &  $10 \le t_{y} \le 14$ & $11 \le t_y \le 15$ \\
\hline
$C_{\cO}^{FLF}/C_{PP}^{FF}C_{PP}^{FF}$ &
$34 \le t_{x} \le 38$& $33 \le t_{x} \le 37$ \\
  &  $10 \le t_{y} \le 14$ & $11 \le t_y \le 15$\\
\hline
$C_{PA}^{FL}(t_x)/C_{PP}^{FF}(t_x)$
& $15 \le t_x \le 23$ & $13 \le t_x \le 23$ \\
\hline
$C_{PP}^{FF}$ & $13 \le t_x \le 23$ & $11 \le t_x \le 23$ \\
\hline
$C_{PP}^{FL}$ & $13 \le t_x \le 23$ & $11 \le t_x \le 23$ \\
\hline
$C_{pp}^{FL}$ & $10 \le t_x \le 20$ & $5 \le t_x \le 23$ \\
\hline
$C_{pa}^{FL}(t_x)/C_{pp}^{FF}(t_x)$ & $15 \le t_x \le 23$ 
& $11 \le t_x \le 23$ \\
\hline\hline
\end{tabular}
\end{center}
\end{table}
\begin{table}
\begin{center}
\caption{\label{tab:spectrum} Lattice spacings, critical hopping
parameters and bare, strange quark masses, obtained as detailed in the
text.  Errors are statistical.}
\begin{tabular}{ccc}
\\
\hline\hline
$\beta$ & 6.2 & 6.0\\ 
\hline
$a^{-1}$ (GeV) & $2.73\err{10}{8}$  & $2.00\err44$\\
$\kappa_{cr}$ & 0.13792(2) & 0.13921(1)\\
$a\tilde m_s$ & 0.0280(19) & 0.0401(17)\\
\hline\hline
\end{tabular}
\end{center}
\end{table}
\begin{table}
\begin{center}
\caption{\label{tab:hqfitparams} 
Results for the fit parameters for the 
heavy-quark-mass dependence of the various quantities studied.
$C_{X}=0$ indicates a linear fit and $C_X=B_X=0$, a fit to a constant.}
\begin{tabular}{ccccc}
\\
\hline\hline
 & $\beta$ & $A_{X}$ & $B_{X}$ & $C_{X}$\\
\hline
$\Phi_B^s(M_{P_l})$ & 6.2 & 0.94(2) & -0.08(1) & 0\\
                    & 6.0 & 0.95(2) & -0.10(2) & 0\\
\hline
$\Phi_B^l(M_{P_l})$ & 6.2 & 0.96(4) & -0.10(2) & 0\\
                    & 6.0 & 0.93(5) & -0.12(3) & 0\\
\hline
$B_{P_s}/B_{P_d}$ & 6.2 & 0.97(3) & 0.02(1) & 0\\
                  & 6.0 & 1.01(4) & 0.03(2) & 0\\
\hline
$\Phi_{\Delta F = 2}^s(M_{P_l})$ & 6.2 & 0.010(1) & -0.29(1) & 0\\
                                 & 6.0 & 0.058(4)& -0.85(2) & 0.22(1)\\
\hline
$\Phi_{\Delta F = 2}^l(M_{P_l})$ & 6.2 & 0.006(1) & -0.29(2) & 0\\
                                 & 6.0 & 0.041(6) & -0.83(6)& 0.20(3)\\
\hline
$(\la\bar P_s|\cO_s^{\Delta F=2}| P_s\ra/
\la\bar P_d|\cO_d^{\Delta F=2}| P_d\ra)$ 
& 6.2 & 1.57(26) & 0.01(6) & 0\\
$\times (M_{P_d}/M_{P_s})$  
& 6.0 & 1.28(14) & 0.11(8) & 0\\  
\hline
$\Phi_f^s(M_{P_l})$ & 6.2 & 0.10(1) & -0.32(4) & 0.04(1)\\
                      & 6.0 & 0.19(1) & -0.53(2) & 0.11(1)\\
\hline
$\Phi_f^l(M_{P_l})$ & 6.2 & 0.09(2) & -0.30(8) & 0.04(2)\\
                      & 6.0 & 0.17(1) & -0.51(4) & 0.11(2)\\
\hline
$(f_{P_s}/f_{P_l})$ & 6.2 & 1.16(2) & 0 & 0\\
$\times\sqrt{M_{P_s}/M_{P_l}}$        & 6.0 & 1.15(1) & 0 & 0\\
\hline\hline
\end{tabular}
\end{center}
\end{table}
\begin{table}
\begin{center}
\caption{\label{tab:results} Results at the two values of the
lattice spacing. The first error on each quantity is statistical while
the others correspond, respectively, to the variations in procedure
described in the first five subsections in \sec{sec:results}.  The
errors enclosed in brackets reflect the variations considered in
quantifying discretization errors. }
\begin{tabular}{ccc}
\\
\hline\hline
$\beta$ & 6.2 & 6.0\\
\hline
$f_{B}$ [MeV] &
177\err{17}{16}$\l[\err{8}{5}\err{11}{0}\r]$
\err{6}{3}\err{0}{6}\err{16}{16}\err{0}{0}
 & 
205\err{9}{8}$\l[\err{14}{9}\err{23}{0}\r]$
\err{9}{4}\err{0}{11}\err{18}{18}\err{0}{0}
\\ 
$\frac{f_{B}}{f_{D_s}}$ &
0.71\err{6}{6}$\l[\err{2}{1}\err{3}{0}\r]$\err{0}{0}\err{0}{2}\err{3}{3}
\err{0}{2}
&
0.76\err{3}{3}$\l[\err{2}{1}\err{6}{0}\r]$\err{0}{0}\err{0}{4}\err{3}{3}
\err{0}{2}\\
$f_{B_s}$ [MeV] &
204\err{12}{12}$\l[\err{10}{6}\err{13}{0}\r]$
\err{8}{3}\err{0}{7}\err{14}{14}\err{6}{0}
&
233\err{5}{5}$\l[\err{17}{11}\err{27}{0}\r]$
\err{10}{4}\err{0}{13}\err{16}{16}\err{6}{0}
\\ 
$\frac{f_{B_s}}{f_{D_s}}$ &
0.82\err{3}{3}$\l[\err{2}{1}\err{4}{0}\r]$\err{0}{0}\err{0}{3}\err{2}{2}
\err{0}{0}&
0.87\err{1}{1}$\l[\err{3}{1}\err{7}{0}\r]$\err{0}{0}\err{0}{5}\err{2}{2}
\err{0}{0}\\
$\frac{f_{B_s}}{f_{B}}$ &
1.15\err{2}{2}$\l[\err{0}{0}\err{0}{0}\r]$\err{0}{0}\err{0}{0}\err{3}{2}
\err{3}{0}&
1.14\err{1}{1}$\l[\err{0}{0}\err{0}{0}\r]$\err{0}{0}\err{0}{0}\err{2}{2}
\err{3}{0}\\
$f_D$ [MeV] &
210\err{10}{9}$\l[\err{5}{4}\err{4}{0}\r]$
\err{8}{3}\err{0}{0}\err{14}{15}\err{0}{0}
&
228\err{4}{4}$\l[\err{7}{7}\err{8}{0}\r]$
\err{10}{4}\err{0}{0}\err{14}{15}\err{0}{0}
\\ 
$f_{D_s}$ [MeV] &
236\err{8}{8}$\l[\err{6}{5}\err{4}{0}\r]$
\err{9}{3}\err{0}{0}\err{12}{12}\err{7}{0}
&
254\err{3}{3}$\l[\err{9}{8}\err{9}{0}\r]$
\err{11}{5}\err{0}{0}\err{11}{12}\err{7}{0}
\\ 
$\frac{f_{D_s}}{f_D}$ &
1.13\err{2}{2}$\l[\err{0}{0}\err{0}{0}\r]$\err{0}{0}\err{0}{0}\err{3}{2}
\err{3}{0}&
1.11\err{1}{1}$\l[\err{0}{0}\err{0}{0}\r]$\err{0}{0}\err{0}{0}\err{2}{2}
\err{3}{0}\\
\hline\hline
\end{tabular}
\end{center}
\end{table}
\begin{table}
\begin{center}
\caption{\label{tab:resultscont} Results at the two values of the
lattice spacing (continued). $r^{\mathrm{direct}}_{sd} = \la\bar
B_s|\cO^{\Delta B=2}_s|B_s\ra/\la\bar
B_d|\cO^{\Delta B=2}_d|B_d\ra$.  An error of $x$
indicates that the variation has not been explicitly performed, but
that it is believed to be small. An error of $y$ means that the
variation has not been explicitely performed.}
\begin{tabular}{ccc}
\\
\hline\hline
$\beta$ 
& 6.2 
& 6.0\\
\hline
$B_{B_d}(M_{B})$ 
& 0.91\err{3}{4}$\l[\err{x}{x}\err{0}{0}\r]$\err{4}{0}\err{x}{x}\err{0}{0}
\err{0}{0}
& 0.89\err{4}{4}$\l[\err{x}{x}\err{0}{0}\r]$\err{3}{0}\err{0}{0}\err{0}{0}
\err{0}{0}
\\
$B_{B_s}(M_{B})$ 
& 0.90\err{2}{2}$\l[\err{x}{x}\err{0}{0}\r]$\err{3}{0}\err{x}{x}\err{0}{0}
\err{0}{0}
& 0.91\err{2}{2}$\l[\err{x}{x}\err{0}{0}\r]$\err{3}{0}\err{0}{0}\err{1}{1}
\err{0}{0}
\\
$B_{D}(M_{D})$ 
& 0.82\err{3}{3}$\l[\err{x}{x}\err{0}{0}\r]$\err{3}{0}\err{x}{x}\err{1}{1}
\err{0}{0}
& 0.81\err{3}{3}$\l[\err{x}{x}\err{0}{0}\r]$\err{2}{0}\err{0}{0}\err{1}{1}
\err{0}{0}
\\
$\frac{B_{B_s}}{B_{B_d}}$ 
& 0.98\err{2}{2}$\l[\err{x}{x}\err{0}{0}\r]$\err{0}{2}\err{x}{x}\err{0}{0}
\err{0}{0}
& 1.02\err{3}{3}$\l[\err{x}{x}\err{0}{0}\r]$\err{0}{0}\err{0}{0}\err{0}{0}
\err{0}{0}
\\
$r^{\mathrm{direct}}_{sd}$ 
& 1.61\err{20}{20}$\l[\err{x}{x}\err{0}{0}\r]$\err{10}{0}\err{y}{x}\err{9}{7}
\err{12}{0}
& 1.36\err{11}{12}$\l[\err{x}{x}\err{0}{0}\r]$\err{1}{0}\err{6}{0}\err{8}{0}
\err{7}{0}\\
\hline\hline
\end{tabular}
\end{center}
\end{table}
\begin{table}
\begin{center}
\caption{\label{tab:resultscont2} Results at the two values of
the lattice spacing (continued). $r^{\mathrm{indirect}}_{sd}=
\l(\frac{M_{B_s}}{M_{B_d}}\r)^2
\l[\frac{f_{B_s}^2B_{B_s}}{f_{B_d}^2B_{B_d}}\r]$.}
\begin{tabular}{ccc}
\\
\hline\hline
$\beta$ 
& 6.2 
& 6.0\\
\hline
$f_{B_d}\sqrt{\hat B_{B_d}^\nlo}$ [MeV] 
& 210\err{20}{21}$\l[\err{10}{6}\err{13}{0}\r]$
\err{12}{0}\err{0}{7}\err{19}{18}\err{0}{0}
& 240\err{12}{11}$\l[\err{16}{11}\err{27}{0}\r]$
\err{15}{5}\err{0}{13}\err{21}{21}\err{0}{0}
\\
$\frac{f_{B_d}}{f_{D_s}}\sqrt{\hat B_{B_d}^\nlo}$ 
& 0.89\err{7}{7}$\l[\err{2}{1}\err{4}{0}\r]$\err{2}{0}\err{0}{3}\err{3}{4}
\err{0}{2}
& 0.95\err{5}{4}$\l[\err{3}{1}\err{7}{0}\r]$\err{2}{0}\err{0}{5}\err{4}{4}
\err{0}{2}
\\
$f_{B_s}\sqrt{\hat B_{B_s}^\nlo}$ [MeV] 
& 241\err{14}{14}$\l[\err{12}{7}\err{15}{0}\r]$
\err{13}{1}\err{0}{9}\err{17}{16}\err{7}{0}
& 277\err{7}{6}$\l[\err{20}{13}\err{32}{0}\r]$
\err{18}{5}\err{0}{15}\err{18}{18}\err{8}{0}
\\
$\frac{f_{B_s}}{f_{D_s}}\sqrt{\hat B_{B_s}^\nlo}$ 
& 1.02\err{4}{4}$\l[\err{3}{1}\err{5}{0}\r]$\err{2}{0}\err{0}{4}\err{2}{2}
\err{0}{0}
& 1.09\err{2}{2}$\l[\err{4}{2}\err{9}{0}\r]$\err{2}{0}\err{0}{6}\err{2}{2}
\err{0}{0}
\\
$f_{D}\sqrt{\hat B_{D}^\nlo}$ [MeV] 
& 222\err{10}{10}$\l[\err{5}{4}\err{5}{0}\r]$
\err{13}{0}\err{0}{0}\err{14}{14}\err{0}{0}
& 240\err{7}{6}$\l[\err{8}{7}\err{8}{0}\r]$
\err{13}{3}\err{0}{0}\err{14}{14}\err{0}{0}
\\
$\frac{f_{D}}{f_{D_s}}\sqrt{\hat B_{D}^\nlo}$ 
& 0.94\err{3}{3}$\l[\err{0}{0}\err{1}{0}\r]$\err{2}{0}\err{0}{0}\err{1}{1}
\err{0}{3}
& 0.94\err{2}{2}$\l[\err{0}{0}\err{0}{0}\r]$\err{1}{0}\err{0}{0}\err{1}{1}
\err{0}{2}\\
$\xi$ 
& 1.15\err{2}{2}$\l[\err{0}{0}\err{0}{0}\r]$\err{0}{1}\err{0}{0}\err{2}{2}
\err{3}{0}
& 1.16\err{2}{2}$\l[\err{0}{0}\err{0}{0}\r]$\err{0}{0}\err{0}{0}\err{3}{2}
\err{3}{0}\\
$r^{\mathrm{indirect}}_{sd}$ 
& 1.38\err{6}{5}$\l[\err{1}{0}\err{0}{0}\r]$\err{0}{3}\err{0}{1}\err{6}{5}
\err{8}{0}
& 1.39\err{5}{5}$\l[\err{1}{0}\err{0}{0}\r]$\err{0}{0}\err{0}{0}\err{6}{5}
\err{8}{0}\\
\hline\hline
\end{tabular}
\end{center}
\end{table}
%
%
\begin{figure}
\centerline{\begin{tabular}{cc}
\epsfxsize=0.48\textwidth\epsffile{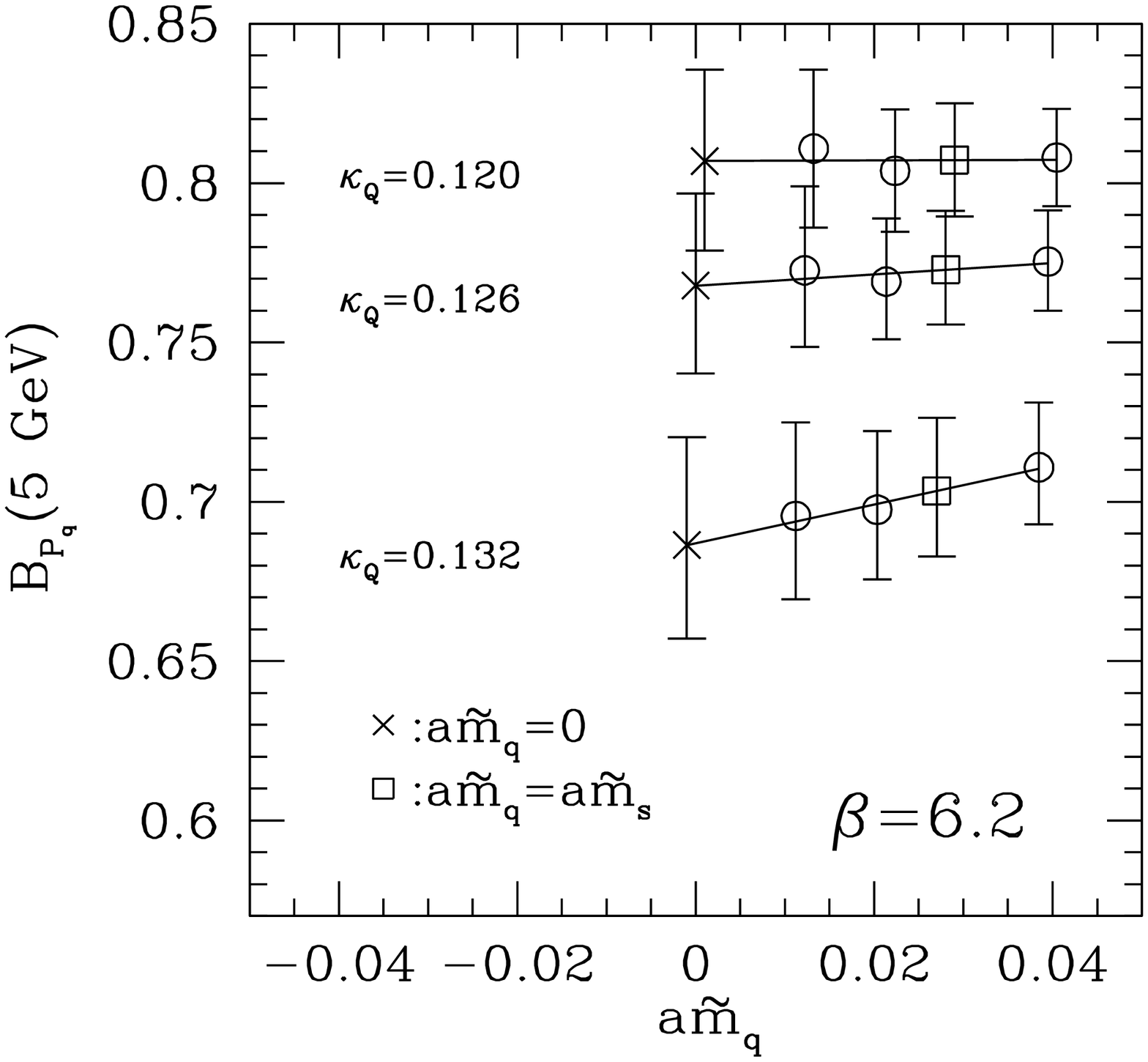}&
\epsfxsize=0.48\textwidth\epsffile{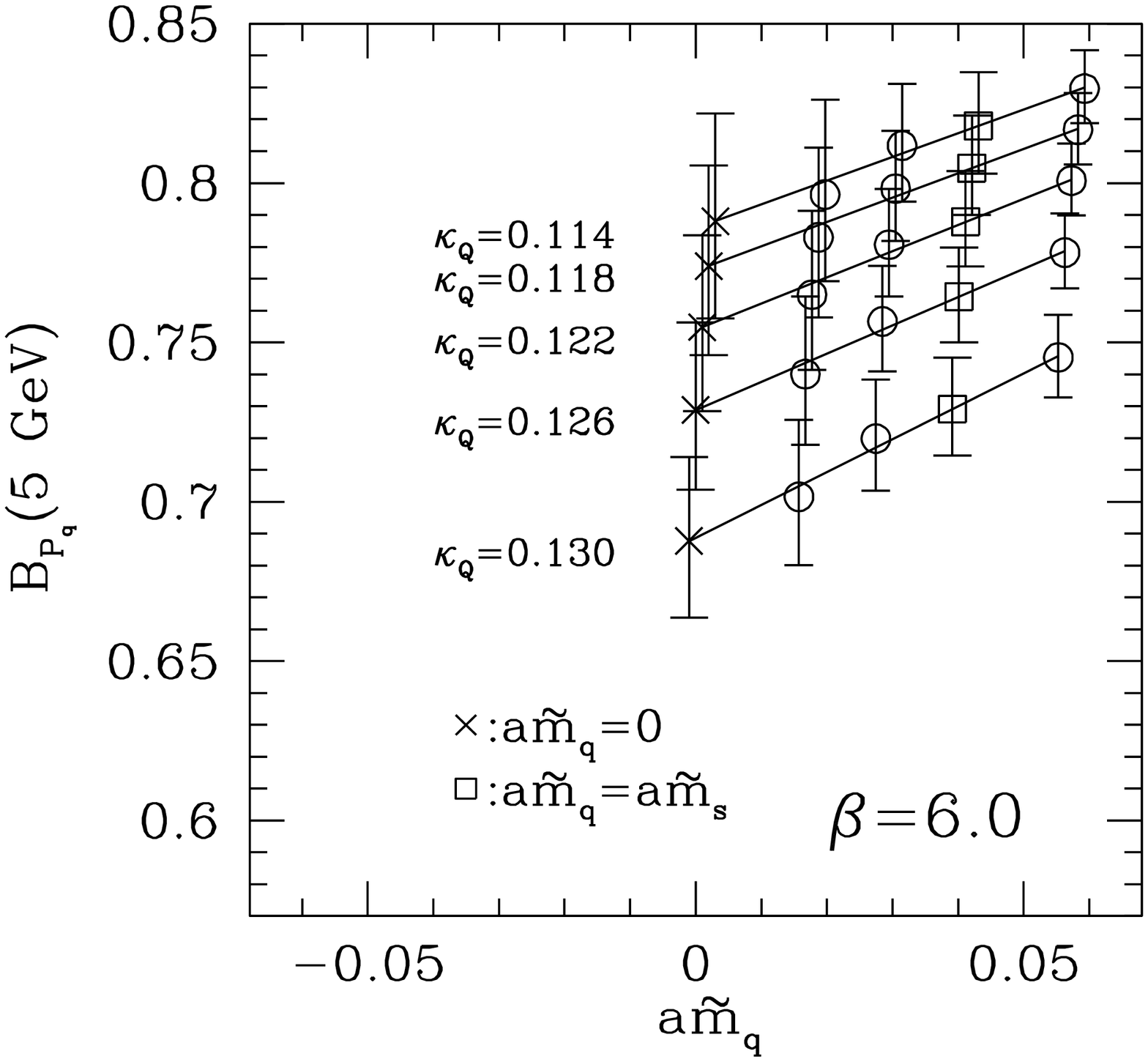}
\end{tabular}}
\vspace{-0.8cm}
\caption{\label{fig:Bvslqm}
Light-quark-mass dependence of the heavy-light $B$-parameter, 
$B_{P_q}(5\gev)$, and extrapolation and interpolation to 
vanishing quark mass and strange-quark mass
at $\beta=6.2$ and 6.0. Points with the same $a\tilde{m}_{q}$ are
shifted for clarity.}
\end{figure}
\begin{figure}
\centerline{\begin{tabular}{cc}
\epsfxsize=0.48\textwidth\epsffile{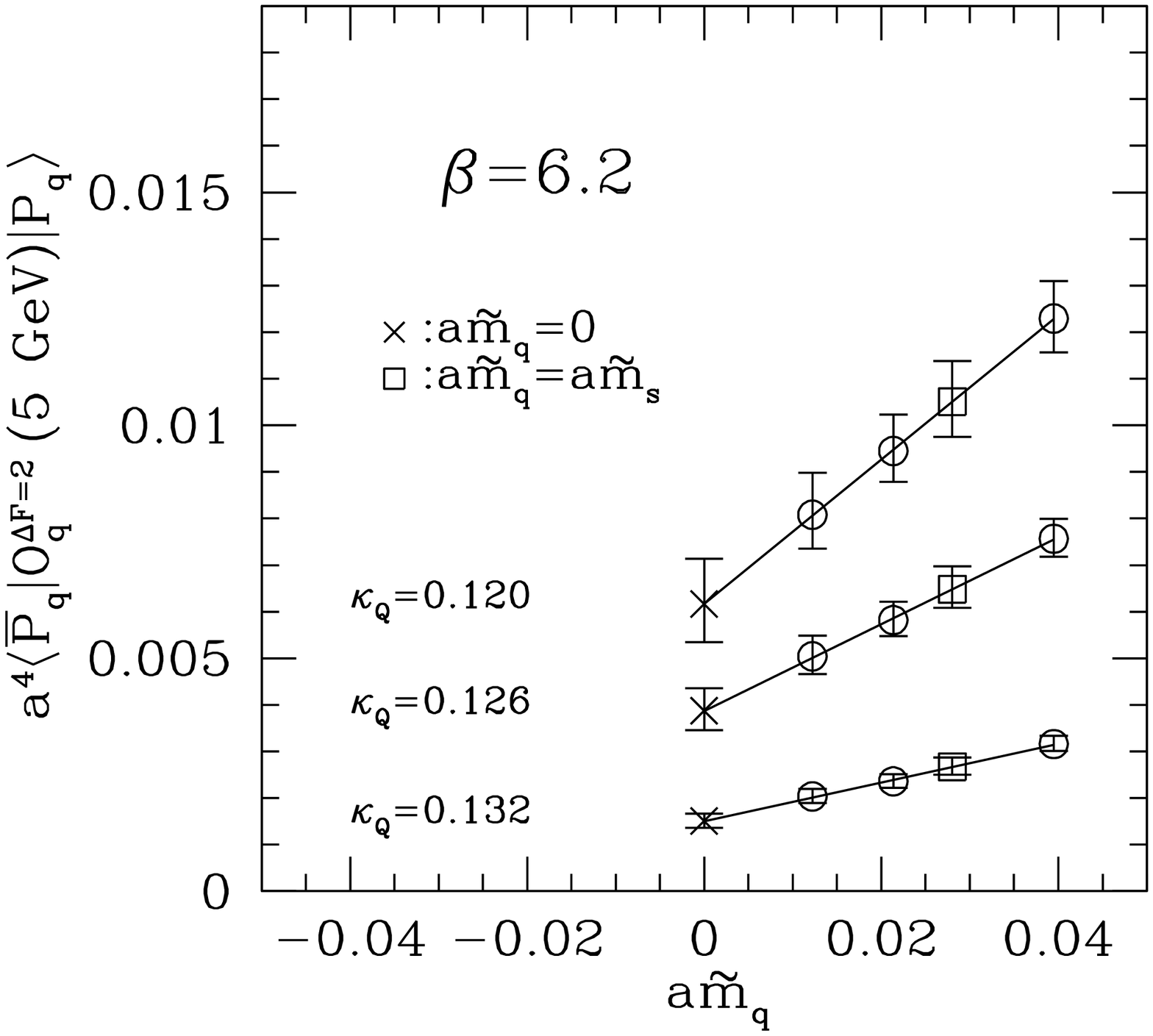}&
\epsfxsize=0.48\textwidth\epsffile{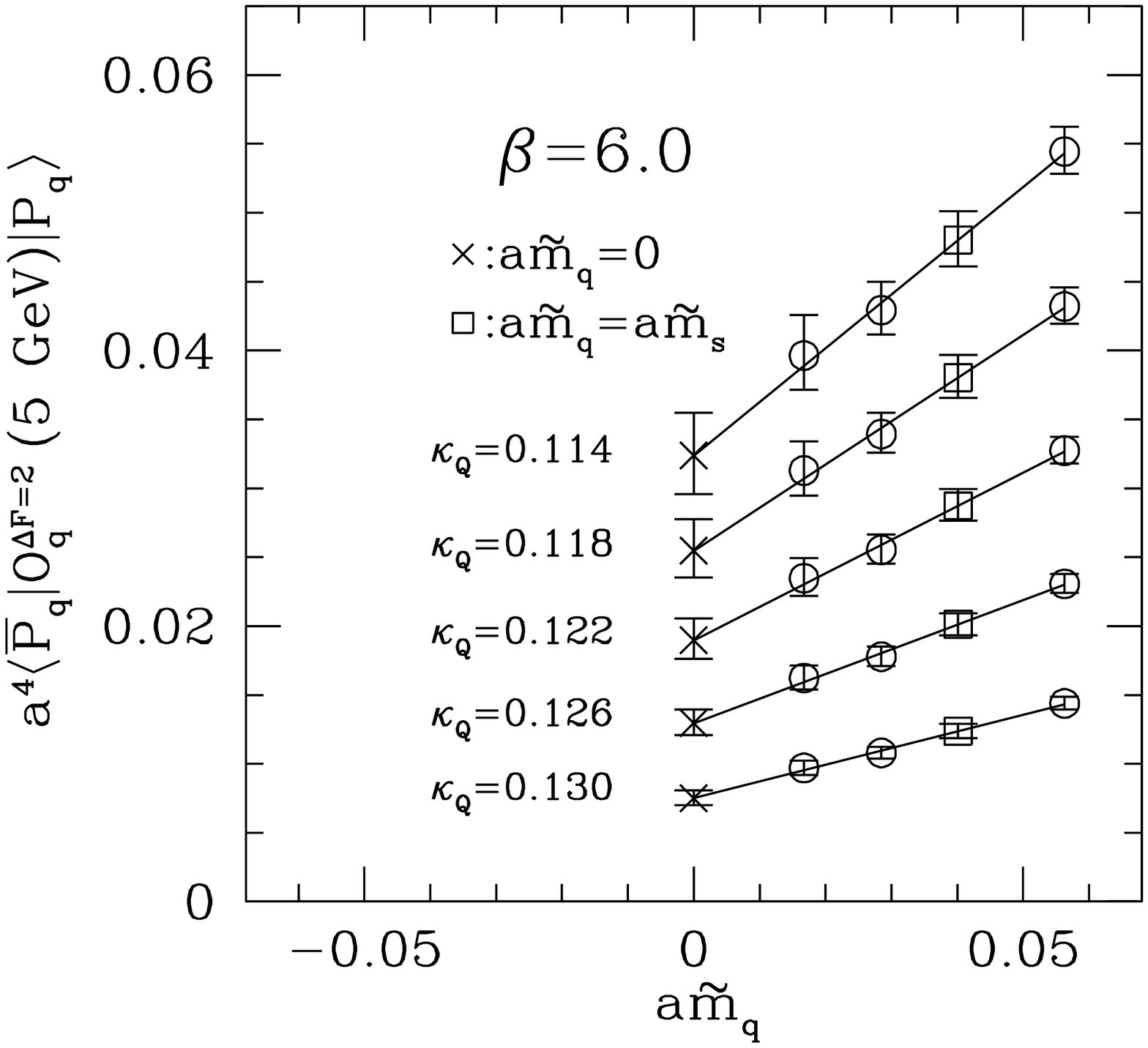}
\end{tabular}}
\vspace{-0.8cm}
\caption{\label{fig:matvslqm}
Light-quark-mass dependence of 
$\la\bar P_q|\cO^{\Delta F=2}_q(5\gev)| P_q\ra$,
and extrapolation and interpolation to 
vanishing quark mass and strange-quark mass
at $\beta=6.2$ and 6.0.}
\end{figure}
\begin{figure}
\centerline{\begin{tabular}{cc}
\epsfxsize=0.48\textwidth\epsffile{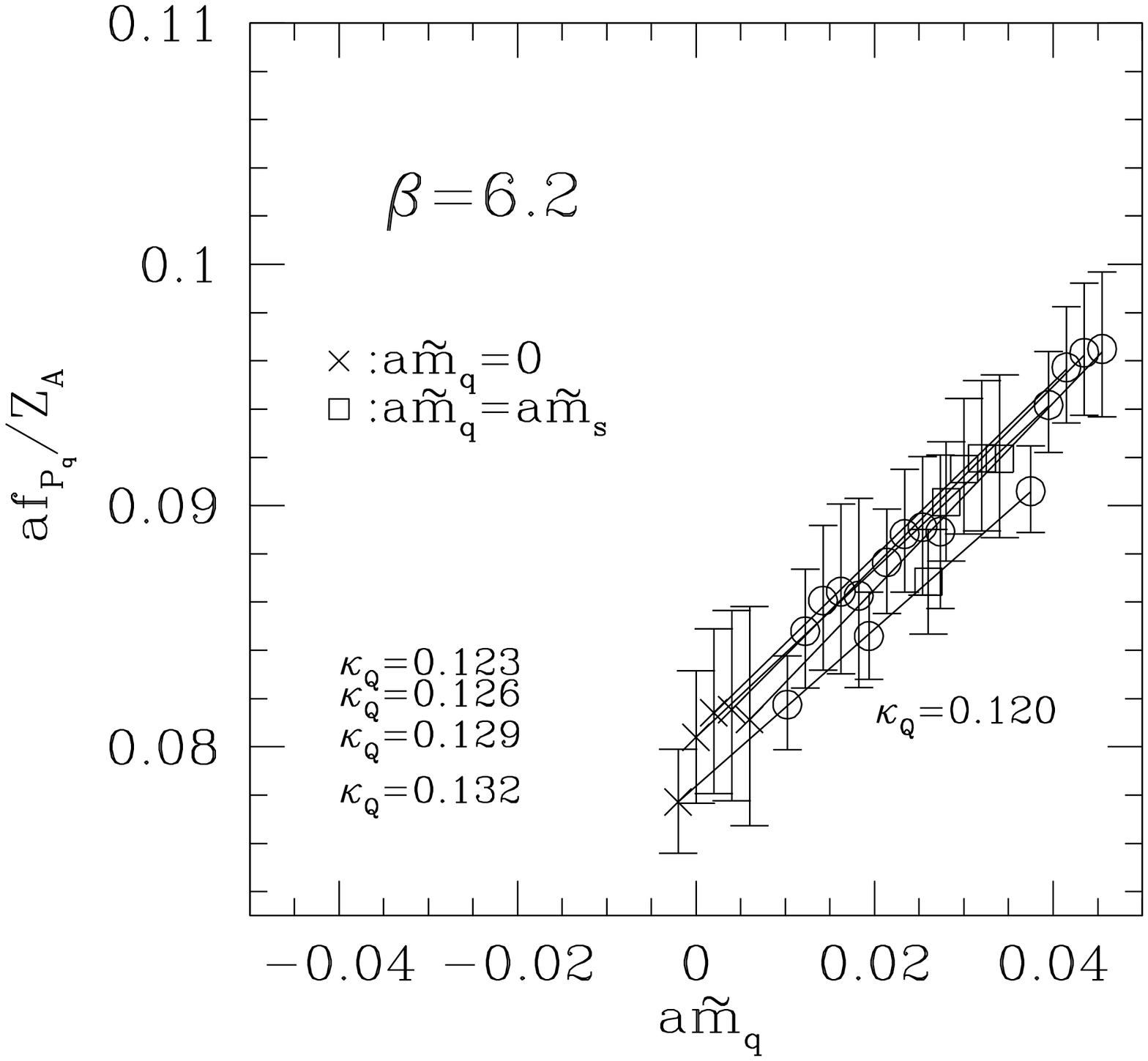}&
\epsfxsize=0.48\textwidth\epsffile{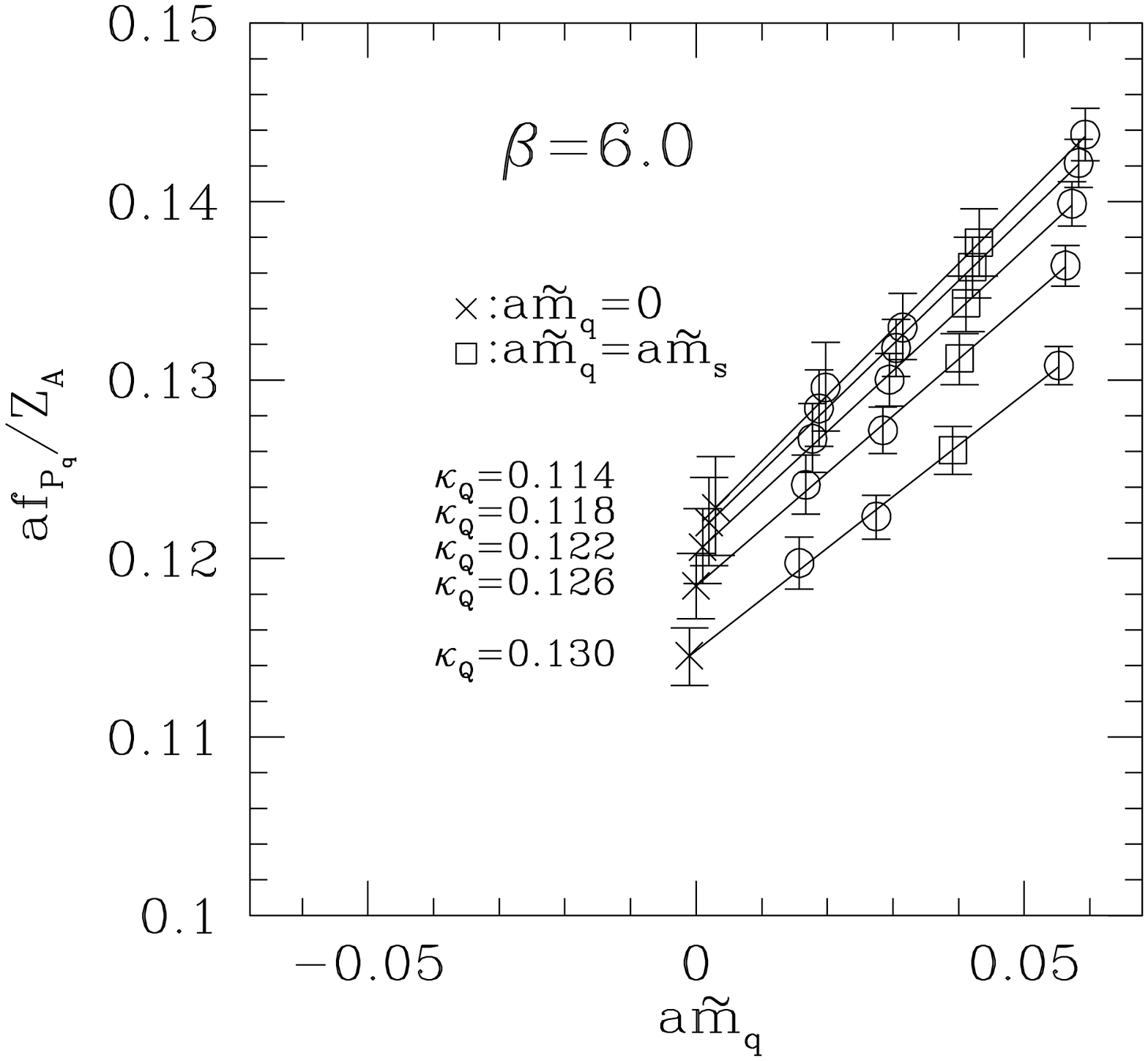}
\end{tabular}}
\vspace{-0.8cm}
\caption{\label{fig:fpvslqm}
Light-quark-mass dependence of the heavy-light decay constant,
$f_{P_q}$, and extrapolation and interpolation to 
vanishing quark mass and strange-quark mass
at $\beta=6.2$ and 6.0.  Points with the same $a\tilde{m}_{q}$ are
shifted for clarity.}
\end{figure}

\end{document}